\documentclass[12pt]{iopart} 
\usepackage{graphicx} 
\usepackage{xcolor}
\usepackage{cite}
\usepackage{amssymb}
\begin{document}

\title{Multidimensional hydrogenic states: Position and momentum expectation values}

\author{J.S. Dehesa$^{1,2}$, D. Puertas-Centeno$^{3}$}

\address{$^1$Departamento de F\'{\i}sica At\'{o}mica, Molecular y Nuclear, Universidad de Granada, 18071-Granada, Spain}
\address{$^2$ Instituto Carlos I de F\'{\i}sica Te\'orica y Computacional, Universidad de Granada, 18071-Granada, Spain}
\address{$^3$Departamento de Matem\'atica Aplicada, Universidad Rey Juan Carlos, 28933 Madrid, Spain}
\ead{dehesa@ugr.es} 

\begin{abstract}
The position and momentum probability densities of a multidimensional quantum system are fully characterized by means of the radial expectation values $\langle r^\alpha \rangle$ and $\left\langle p^\alpha \right\rangle$, respectively. These quantities, which describe and/or are closely related to various fundamental properties of realistic systems, have not been calculated in an analytical and effective manner up until now except for a number of three-dimensional hydrogenic states. In this work we explicitly show these expectation values for all discrete stationary $D$-dimensional hydrogenic states in terms of the dimensionality $D$, the strength of the Coulomb potential (i.e., the nuclear charge) and the $D$ state's hyperquantum numbers. Emphasis is placed on the momentum expectation values (mostly unknown, specially the ones with odd order) which are obtained in a closed compact form. Applications are made to circular, $S$-wave, high-energy (Rydberg) and high-dimensional (pseudo-classical) states of three- and multidimensional hydrogenic atoms. This has been possible because of the analytical algebraic and asymptotical properties of the special functions (orthogonal polynomials, hyperspherical harmonics) which control the states' wavefunctions. Finally, some Heisenberg-like uncertainty inequalities satisfied by these dispersion quantities are also given and discussed.
\end{abstract}


\maketitle

\section{Introduction}

Space dimensionality $D$ is a fundamental variable of quantum single and many body systems which substantially modifies the physical solutions of their wave equations \cite{Dong2011,Yanez1994,Aquilanti1997,Avery2006}, and consequently all their chemical and physical properties. The theoretical analysis of these properties in $D$-dimensional quantum physics leans heavily on accurate knowledge of the radial expectation values of the position and momentum probability densities $\rho(\vec{r})$ and $\gamma(\vec{p})$, respectively. These quantities are given by $ \langle r^{\alpha} \rangle = \int_{\mathbb{R}_D} r^{\alpha} \rho(\vec{r}) d \vec{r}$ and $ \langle p^{\alpha} \rangle = \int_{\mathbb{R}_D} p^{\alpha} \gamma(\vec{p})\,d \vec{p}$, respectively, where $r = |\vec{r}|$ and $p = |\vec{p}|$. They are physically meaningful for specific values of the order $\alpha$ and, when $D=3$, they are experimentally accessible since the seventies by positron annihilation, binary (e, 2e) or electron momentum spectroscopy and inelastic (Compton) scattering of X-rays, $\gamma$ rays, and high-energy electrons \cite{Bonham1974,Williams1977,Glusker1986,Coppens1997,Weigold1999,Thakkar2004,Thakkar2005,Blair2014,Hart2005,Thakkar1990,Thakkar1980,Lehtola2013}. Moreover they describe, save for a proportionality factor, relevant quantities \cite{Williams1977,Gadre1991,March1987,Thakkar2004,Thakkar2005,Blair2014,Delbourgo2009} such as e.g., the electron-nucleus attraction energy ($\langle r^{-1}\rangle$), the diamagnetic susceptibility ($\langle r^{2}\rangle$), the Dirac-Slater exchange energy ($\langle p\rangle$), the electron kinetic energy ($\langle p^{2}\rangle$), the interelectronic repulsion energy ($\langle p^{3}\rangle$), the Breit-Pauli relativistic correction to the kinetic energy ($\langle p^{4}\rangle$), the height of the peak of the Compton profile within the impulse approximation and the energy level change induced by the Born reciprocity effect on all hydrogenic bound state levels relevant at atomic and galactic scales ($\langle p^{-1}\rangle$). \\

Nowadays much attention is being centered around the main prototype of multidimensional many-body systems, the $D$-dimensional hydrogenic system, because of its enormous interest in quantum chemistry and atomic and molecular physics \cite{Nieto1979,Herschbach1993,Tsipis1996,Herschbach1996,Aquilanti2001,Dehesa2011,Herschbach2020,Aquilanti1997,Coletti2013,Harrison2005,Dehesa2010,Dulieu2018}, classical and quantum information \cite{Yanez1994,Dehesa2010,Dehesa2011,Toranzo2016,Toranzo2017b,Puertas2018a,Witten2018}, high energy physics \cite{Witten1980,Hoof2016,Bures2015,Corda2018} and electronic complexity \cite{Dehesa2011,Sen2012,Lopez2013,Sobrino-coll2017}, among many other chemical and physical fields. This system is the main prototype to model numerous properties and phenomena of multidimensional quantum many-body systems with standard ($D = 3$) and non-standard ($D\neq 3$) dimensionalities \cite{Witten1980,Herschbach1993,Burgbacher1999,Andrew1990}. It includes a large variety of three-dimensional physical systems (e.g., hydrogenic atoms and ions, exotic atoms, antimatter atoms, Rydberg atoms) and a number of nanotechnological objects which have been shown to be very useful in semiconductor nanostructures (e.g., quantum wells, wires and dots) \cite{Harrison2005,Li2007}. \\ 

Although tremendous advances have been witnessed in understanding the position properties of the multidimensional quantum systems, the momentum properties are not so well known in spite of the efforts of many authors (see e.g. \cite{Gadre1991,Aquilanti2001,Thakkar2004,Thakkar2005,Dehesa2010,Dehesa2011,Coletti2013,Delbourgo2009,Pathak1981}). This is because the required computational work is much bigger and the necessary experiments are less accessible. So, the analytical determination of these properties are very welcome; however, this is only possible in a reduced number of multidimensional quantum systems such as those of harmonic and hydrogenic character.\\

 The aim of this work is to cover this lack by means of the analytical determination of the momentum expectation values of any order for the $D$-dimensional hydrogenic systems. Indeed, we will determine in a closed and simple form the momentum expectation values of arbitrary order for all the discrete stationary states of $D$-dimensional quantum systems of hydrogenic character in terms of the strength of the Coulomb potential (i.e., the nuclear charge $Z$), the dimensionality and the  $D$ hyperquantum numbers $(\eta,\mu_1,\mu_2,\ldots,\mu_{D-1})$ which characterize the quantum state. \\

The structure of the paper is the following. In Section \ref{sec:probdensities} we briefly describe the position and momentum wavefunctions and the associated probability densities for all discrete stationary multidimensional hydrogenic states. In Section \ref{sec:expectvalues} we give the analytical expressions for the radial expectation values in terms of the potential parameters and the state's hyperquantum numbers by means of generalized hypergeometric functions of $_pF_q(1)$ type. These expressions, however, work efficiently for the position expectation values of any order and for the momentum expectation values of even order only. The expectation values of the  extreme high-energy and high-dimensional states are also considered; they require a very different approach, based on the asymptotics of the Laguerre and Gegenbauer orthogonal polynomials. Section \ref{sec:oddmomvalues} is monographically devoted to the momentum expectation values of real order for the multidimensional hydrogenic system, with emphasis on those of odd order, which are applied for illustration to two mostly unknown and interesting cases: the mean momentum and the average inverse momentum. Applications to three-dimensional hydrogenic atoms and to relevant multidimensional circular and ($nS$)-wave states are also done. In Section \ref{sec:posmomur} we give various formalizations of the position-momentum Heisenberg principle of multidimensional quantum physics based on uncertainty inequalities which connect the momentum expectation values with the position expectation values or with the position entropic moments of multidimensional hydrogenic systems \cite{Nagy1999,Liu1996,Liu1997,Angulo2000} . Finally, some conclusions and three appendices are given.

\section{The multidimensional hydrogenic probability densities}
\label{sec:probdensities}
In this section we briefly describe the wavefunctions and the associated probability densities for the discrete stationary states of the $D$-dimensional hydrogenic system in both position and momentum spaces. This system is composed by a negatively-charged particle moving around a positively charged core under the Coulomb potential $V_{D}(r) = - \frac{Z}{r}$, where $r=|\vec{r}|$ and $Z$ is the charge of the nuclear core, assumed to be pointwise and located at the origin. Later on, we give the associated probability densities for the quantum states of the system. Atomic units (i.e., $\hbar = m_e = e = 1$) are used throughout the paper. \\
The momentum wavefunctions can be obtained by means of the Fourier transform of the  position wavefunctions of the system, which are the known physical solutions of the corresponding Schr\"{o}dinger equation 
\begin{equation}\label{eqI_cap1:ec_schrodinger}
\left( -\frac{1}{2} \vec{\nabla}^{2}_{D} - \frac{Z}{r}\right) \Psi \left( \vec{r} \right) = E \Psi \left(\vec{r} \right),
\end{equation}
where $\vec{\nabla}_{D}$ denotes the $D$-dimensional gradient operator, and the  position vector $\vec{r}  =  (x_1 ,  \ldots  , x_D)$ in hyperspherical units  is  given as $(r,\theta_1,\theta_2,\ldots,\theta_{D-1})      \equiv
(r,\Omega_{D-1})$, where $r \equiv |\vec{r}| = \sqrt{\sum_{i=1}^D x_i^2}
\in [0 , \: +\infty)$  and $x_i =  r \left(\prod_{k=1}^{i-1}  \sin \theta_k
\right) \cos \theta_i$ for $1 \le i \le D$
and with $\theta_i \in [0 , \pi), i < D-1$, $\theta_{D-1} \equiv \phi \in [0 ,  2
\pi)$. 

\subsection{The wavefunctions}

 The position wavefunctions are characterized \cite{Yanez1994,Dehesa2010} by the energetic eigenvalues known as 
 \begin{equation}
E =-\frac{Z^{2}}{2\eta^{2}}; \quad \eta = n+ \frac{D-3}{2}, \quad n=1,2,3,\ldots,
\end{equation}
where $\eta$ is the principal hyperquantum number associated to the radial variable $r$, and the associated eigenfunctions expressed as
\begin{equation}\label{eq:FunOndaPos}
\Psi_{n,l, \left\lbrace \mu \right\rbrace }(\vec{r})=\mathcal{R}_{n,l}(r) \times {\cal{Y}}_{l,\{\mu\}}(\Omega_{D-1}),
\end{equation}
where $(l,\left\lbrace \mu \right\rbrace)\equiv(l\equiv\mu_1,\mu_2,\ldots,\mu_{D-1})$ denote the hyperquantum numbers associated to the angular variables $\Omega_{D-1}\equiv (\theta_1, \theta_2,\ldots,\theta_{D-1})$, which have the values  $l= 0,1,2, \ldots, n-1$ and $l\geq \mu_{2} \geq \ldots \geq \mu_{D-1} \equiv |m | \geq 0$. The angular part of the eigenfunctions are the hyperspherical harmonics, $\mathcal{Y}_{l,\{\mu\}}(\Omega_{D-1})$, defined \cite{Yanez1994,Aquilanti2001,Coletti2013,Avery2006} as
\begin{equation}
\label{eq:24}
\mathcal{Y}_{l,\{\mu \}}(\Omega_{D-1}) =\frac{1}{\sqrt{2\pi}}e^{im\phi}\prod_{j=1}^{D-2}\tilde{C}^{(\alpha_{j}+\mu_{j+1})}_{\mu_{j}-\mu_{j+1}}(\cos\theta_{j}) (\sin\theta_{j})^{\mu_{j+1}},
\end{equation}
where $2\alpha_{j} = D-j-1$, $\widetilde{C}_m^{(\alpha)}(t)$ denotes the orthonormal Gegenbauer or ultraspherical polynomial \cite{Olver2010} of degree $m$ and parameter $\alpha$, and with the values $0 \leq \theta_j \leq \pi$ ($j=1,2,\dots,D-2$) and $0 \leq \phi \leq 2\pi$.
Moreover, these hyperfunctions satisfy the orthonomalization condition
\begin{equation}\label{eq:normalizacionY}
\int_{\mathcal{S}_{D-1}} {\cal{Y}}^{*}_{l', \left\lbrace \mu' \right\rbrace}(\Omega_{D-1})
{\cal{Y}}_{l, \left\lbrace \mu \right\rbrace}\left( \Omega_{D-1} \right) d\Omega_{D-1 } = \delta_{l,l'} \delta_{\left\lbrace \mu \right\rbrace,
\left\lbrace \mu' \right\rbrace}
\end{equation}
%
Moreover, the radial part of the eigenfunctions $\mathcal{R}_{n,l}(r)$ are known to have the form \cite{Yanez1994}
\begin{eqnarray}
\mathcal{R}_{n,l}(r) &=& K_{n,l}\left(\frac{2Z}{\eta}r\right)^{l}e^{-\frac{Z}{\eta}r}\mathcal{L}_{n-l-1}^{(2l+D-2)}\left(\frac{2Z}{\eta}r\right)\nonumber\\
&=& K_{n,l}\left[\frac{\omega_{2L+1(\tilde{r})}}{\tilde{r}^{D-2}}\right]^{\frac{1}{2}} \mathcal{L}_{\eta-L-1}^{(2L+1)}(\tilde{r})\nonumber\\
&=& \left(\frac{2Z}{\eta}\right)^{\frac{D}{2}} \left(\frac{1}{2\eta}\right)^{1/2}\left[\frac{\omega_{2L+1(\tilde{r})}}{\tilde{r}^{D-2}}\right]^{1/2}\tilde{\mathcal{L}}_{\eta-L-1}^{(2L+1)}(\tilde{r})
\end{eqnarray}
where $L=l+\frac{D-3}{2}$ (so that $2L+1=2l+D-2$), and $\widetilde{r}=\frac{2Z}{\eta}r$. The 
symbols  $\mathcal{L}^{(\alpha)}_{k}(x)$ y $\tilde{\mathcal{L}}^{(\alpha)}_{k}(x)$ denote the orthogonal and orthonormal Laguerre polynomials \cite{Olver2010} with respect to the weight $\omega_{\alpha}(x)=x^\alpha e^{-x}$ on the interval $\left[0,\infty\right)$, respectively, which are related by  
\begin{equation}
\tilde{\mathcal{L}}^{(\lambda)}_{k}(x)=\left[ \frac{k!}{\Gamma(k+\lambda+1)} \right]^\frac{1}{2} \mathcal{L}^{(\lambda)}_{k}(x),
\end{equation}
and the normalization constant $K_{n,l}$ is given by
\begin{eqnarray}
K_{n,l} &=& \left(\frac{2Z}{\eta}\right)^{\frac{D}{2}}\left\{\frac{(\eta-L-1)!}{2\eta(\eta+L)!}\right\}^{\frac{1}{2}}\nonumber\\
 &=& \left\{\left(\frac{2Z}{n+\frac{D-3}{2}}\right)^{D}\frac{(n-l-1)!}{2(n+\frac{D-3}{2})(n+l+D-3)!}\right\}^{\frac{1}{2}},
\end{eqnarray}
which guaranties  $\int_{\mathbb R_D} \left| \Psi_{\eta,l, \left\lbrace \mu \right\rbrace }(\vec{r}) \right|^2 d\vec{r} =1$.  
Remark that the position wavefunction $\Psi_{n,l, \left\lbrace \mu \right\rbrace }(\vec{r})$ is indeed normalized to unity since the hyperspherical harmonics given by $\int_{\Omega_{D}} |\mathcal{Y}_{l,\{\mu \}}(\Omega_{D})|^{2}d\Omega_{D} = 1$.
\\

Then, the eigenfunctions for a generic stationary state $(n,l,\{\mu\})$ of the $D$-dimensional hydrogenic system in momentum space is obtained by means of the corresponding Fourier transform of the position eigenfunctions (\ref{eq:FunOndaPos}), 
\begin{equation}
	\widetilde{\Psi}_{n,l,\{\mu\}}(\vec{p})= \int_{\mathbb{R}_D} e^{-i\vec{p}\cdot \vec{r}}\,\Psi_{n,l, \left\lbrace \mu \right\rbrace }(\vec{r})\,d^D\vec{r},
	\end{equation}
obtaining
\begin{equation}\label{eq:hwavemom}
\widetilde{\Psi}_{n,l,\{\mu\}}(\vec{p})=\mathcal{M}_{n,l} (p) \times {\cal{Y}}_{l,\{\mu\}}(\Omega_{D-1}),
\end{equation}
where $\vec{p} = (p, \theta_{1}, \ldots, \theta_{D-1})$, and the radial momentum wavefunction is 
\begin{eqnarray}
\mathcal{M}_{n,l}(p) &=& K'_{n,l}\frac{(\eta\tilde{p})^{l}}{(1+\eta^{2}\tilde{p}^{2})^{L+2}}\,\, \mathcal{C}_{\eta-L-1}^{(L+1)}\left(\frac{1-\eta^{2}\tilde{p}^{2}}{1+\eta^{2}\tilde{p}^{2}}\right)\nonumber\\
&=& \left(\frac{\eta}{Z}\right)^{D/2}(1+y)^{3/2}\left(\frac{1+y}{1-y}\right)^{\frac{D-2}{4}}\sqrt{\omega_{L+1}^{*}(y)}\,\,\widetilde{\mathcal{C}}^{L+1}_{\eta-L-1}(y),
\end{eqnarray}
with $y=\frac{1-\eta^2\tilde{p}^2}{1-\eta^2\tilde{p}^2}$, $\tilde{p}=\frac{p}{Z}$ and the normalization constant 
\begin{eqnarray}
K'_{n,l} &=& 2^{2L+3}\left(\frac{(\eta-L-1)!}{2\pi(\eta+L)!}\right)^{1/2}\Gamma(L+1)\eta^{\frac{D+1}{2}}\nonumber\\
&=&  2^{2l+D}\left(\frac{(n-l-1)!}{2\pi(n+l+D-3)!}\right)^{1/2}\Gamma(l+\frac{D-1}{2})\,\eta^{\frac{D+1}{2}}
\label{mocte}
\end{eqnarray}
Note that $\eta-L-1=n-l-1$ and $L+1 = l+\frac{D-1}2$. The symbols $\mathcal{C}_m^{(\alpha)}(y)$ and $\widetilde{\mathcal{C}}_m^{(\alpha)}(y)$ denote 
the orthogonal and orthonormal Gegenbauer polynomials \cite{Olver2010} with respect to the weight function $\omega_\alpha^* (y)=(1-y^2)^{\alpha-\frac{1}{2}}$ on the interval $[-1,+1]$, respectively, which are mutually related by 
\begin{equation}
\widetilde{\mathcal{C}}^{(\lambda)}_{k}(x)=\left(\frac{k!(k+\lambda)\Gamma^{2}(\lambda)}{\pi2^{1-2\lambda}\Gamma(2\lambda+k)}\right)^{1/2}\mathcal{C}^{(\lambda)}_{k}(x).
\end{equation}
so that
$$\widetilde{\mathcal{C}}^{(l+\frac{D-1}2)}_{n-l-1}(x)=A(n,l;D)^\frac12\,\,{\mathcal{C}}^{(l+\frac{D-1}2)}_{n-l-1}(x)$$
 with the constant
\begin{equation}
A(n,l;D) = \frac{(n-l-1)!(n+\frac{D-3}{2})[\Gamma(l+\frac{D-1}{2})]^{2}}{2^{2-2l-D}\pi\Gamma(n+l+D-2)}.
\end{equation} 
Let us point out that the position and momentum $D$-dimensional wavefunctions (\ref{eq:FunOndaPos}) and
(\ref{eq:hwavemom}), respectively, reduce to the corresponding three-dimensional wavefunctions obtained by numerous authors (see e.g. \cite{Fock1935,Podolsky1929,Hey1993}).

\subsection{The probability densities}
 
Finally, the corresponding position and momentum probability densities of a $D$-dimensional hydrogenic system are given in terms of the hyperquantum numbers, $(n,l,\{\mu\})$, by 
\begin{eqnarray}
\label{eq:denspos}
\rho_{n,\{\mu\}}(\vec{r}) &=& \left|{\Psi}_{n,l,\{\mu\}} (\vec{r}) \right|^2 =\mathcal{R}_{nl}^{2}(r) \times |\mathcal{Y}_{l,\{\mu\}}(\Omega_{D-1})|^{2}\nonumber \\
&=& K_{n,l}^{2}\tilde{r}^{2l}e^{-\tilde{r}}[\mathcal{L}_{n-l-1}^{(2l+D-2)}(\tilde{r})]^{2} \times |\mathcal{Y}_{l,\{\mu\}}(\Omega_{D-1})|^{2}\nonumber\\
&=& \left(\frac{2Z}{\eta}\right)^{D} \frac{1}{2 \eta} \,\, \frac{\omega_{2L+1}(\tilde{r})}{\tilde{r}^{D-2}}\,\,[{\widehat{\mathcal{L}}}_{\eta-L-1}^{(2L+1)}(\tilde{r})]^{2} \times |\mathcal{Y}_{l,\{\mu\}}(\Omega_{D-1})|^{2}.
\end{eqnarray}
in position space, and
\begin{eqnarray}
\label{eq:hgamma}
\gamma(\vec{p})&=&\left| \widetilde{\Psi}_{n,l,\{\mu\}} (\vec{p}) \right|^2= \mathcal{M}_{n,l}^2(p) \times
\left|{\cal{Y}}_{l,\{\mu\}}(\hat{\Omega}_{D-1})\right|^2\nonumber \\
&=& K_{n,l}'^{2}\frac{(\eta \tilde{p})^{2l}}{(1+\eta^{2}\tilde{p}^{2})^{2L+4}}\left[\mathcal{C}_{\eta-L-1}^{(L+1)}\left( \frac{1-\eta^{2}\tilde{p}^{2}}{1+\eta^{2}\tilde{p}^{2}} \right)\right]^{2}\times|\mathcal{Y}_{l,\{\mu \}}(\Omega_{D-1})|^{2}\nonumber \\
&=& \left(\frac{\eta}{Z} \right)^{D}(1+y)^{3}\left(\frac{1+y}{1-y}\right)^{\frac{D-2}{2}}\omega^{*}_{L+1}(y)\,\,[{\tilde{\cal{C}}}^{(L+1)}_{\eta-L-1}(y)]^{2}\times|\mathcal{Y}_{l,\{\mu \}}(\Omega_{D-1})|^{2}\nonumber \\
\end{eqnarray}
 in momentum space, where $y=\frac{1-\eta^2\tilde{p}^2}{1-\eta^2\tilde{p}^2}$ and $\omega_\alpha^* (y)=(1-y^2)^{\alpha-\frac{1}{2}}$.

\section{The position and momentum expectation values}
\label{sec:expectvalues}
In this section we briefly review the determination of the radial expectation values of the $D$-dimensional hydrogenic system in both position and momentum spaces, $\langle r^\alpha \rangle$ and $\left\langle p^\alpha \right\rangle$, respectively. We show that the position expectation values $\langle r^\alpha \rangle$ are well controlled for all $\alpha$'s, but the approach to find the momentum expectation values $\left\langle p^\alpha \right\rangle$, although valid for any real $\alpha$, is shown to work efficiently for even $\alpha$'s only. Moreover, we point out the position and momentum expectation values for the high-energy (i.e., Rydberg) and the high-dimensional (i.e., pseudo-classical) states, which have been obtained by very different means.   \\

The spreading of these physical probability densities $\rho(\vec{r})$, which control all the
macroscopic physical and chemical properties of the hydrogenic system, is conventionally
measured by means of the power moments or radial expectation values
\begin{eqnarray}
\langle r^\alpha \rangle &=& \int_{\mathbb{R}_D} r^\alpha \rho(\vec{r})d \vec{r} = 
\int_0^\infty r^{\alpha+D-1}R_{n,l}^2(r)dr\nonumber\\
&=& \frac{1}{2\eta}\left(\frac{\eta}{2Z}\right)^{\alpha}\int_{0}^{\infty} \omega_{2l+D-2}(t)\,\,[\tilde{\mathcal{L}}^{(2l+D-2)}_{n-l-1}(t)]^{2}\, t^{\alpha+1}\, d\,t
 \label{eq:radexpec2}
\end{eqnarray}
(which holds for $\alpha > -D-2l$) in position space, and
\begin{eqnarray}
\left\langle p^\alpha \right\rangle &=& \int p^\alpha \gamma(\vec{p}) d \vec{p}= 
\int_0^\infty p^{\alpha+D-1}\mathcal{M}_{n,l}^2(p)dp \nonumber \\
&=& \left( \frac{Z}{\eta} \right)^{\alpha} \mathcal{K}_{n,l}\int_{-1}^{+1}  \left[ {\cal{C}}_{k}^{(\nu)}(t) \right]^2 \left(  1-t \right)^{\nu+\frac{\alpha-1}{2}} \left(  1+t \right)^{\nu-\frac{\alpha-1}{2}}dt
\label{eq:moexpec1}\\
&=& \left( \frac{Z}{\eta} \right)^{\alpha} \int_{-1}^{+1} \omega_{\nu}^{*}(t) \left[ {\tilde{\cal{C}}}_{k}^{(\nu)}(t) \right]^2 \left(  1-t \right)^{\frac{\alpha}{2}} \left(1+t \right)^{1-\frac{\alpha}{2}}dt,
\label{eq:moexpec}
\end{eqnarray}
(which holds for $-2l-D \le\alpha\le 2l+D+2$) in momentum space, respectively. Note that $k = \eta - L -1 = n-l-1$, $\nu = L+1 = l + (D-1)/2$, $\omega_\nu^* (t)=(1-t^2)^{\nu-\frac{1}{2}}=(1-t^2)^{l+\frac{D-2}{2}}$ is the weight function of the Gegenbauer polynomials ${\tilde{\cal{C}}}_{k}^{(\nu)} (t)$, and the constant
\begin{eqnarray} \label{eq:karara}
	\mathcal{K}_{n,l} &=&  \frac{ K_{n,l}'^{2}}{2^{2l+D+1} \eta^{D}} = \, 2^{2(L+1)}\eta\left[\Gamma(L+1)\right]^2 \left(\frac{(\eta-L-1)!}{2\pi(\eta+L)!}\right)
\end{eqnarray}
Moreover, it is important to realize that  both position and momentum expectation values depend on $(\eta,L,Z)$ only; that is, on the principal and orbital hyperquantum numbers $(n,l)$, the dimensionality $D$ and the nuclear charge $Z$ only.\\
From these two expressions one obtains
\begin{eqnarray}
\langle r^{\alpha} \rangle &=& \left(\frac{D-1}{4Z}\right)^{\alpha}\frac{\Gamma (D+\alpha )}{\Gamma (D)}; \quad \alpha > -D  \\
\label{eq:posexpec}
\langle p^{\alpha} \rangle &=& \left(\frac{2Z}{D-1} \right)^{\alpha}\frac{2\,\Gamma(\frac{D-\alpha}{2}+1)\Gamma(\frac{D+\alpha}{2})}{D\,\Gamma^{2}\left(\frac{D}{2} \right)}; \quad -D<\alpha<D+2
\label{eq:momexpec}
\end{eqnarray}
for the position and momentum expectation values of the ground state $(n=1, l=0)$ of the $D$-dimensional hydrogenic state, respectively.\\
Moreover, taking into account Eqs. (\ref{eq:radexpec2}) and (\ref{eq:moexpec}) and the integral representation of the generalized hypergeometric functions ${}_{p+1}F_{p}(1)$ \cite{Olver2010,Slater1966,Luke1969}, it is possible to find 
\begin{eqnarray} \nonumber
	\langle r^\alpha \rangle =\frac{\eta^{\alpha-1}}{2^{\alpha+1}Z^{\alpha}}\frac{\Gamma(2L+\alpha+3)}{\Gamma(2L+2)}\\\label{eq:posexp} 
	\times \,\, _3F_2\left(
\left.
\begin{array}{ll}
-\eta+L+1,&-\alpha-1,\,\alpha+2\\
&2L+2,\,1
\end{array}
\right|
1
\right),
\end{eqnarray}
for the position expectation values \cite{Andrae1997,Drake1990,Tarasov2004,Dehesa2010}, and 
\begin{eqnarray} \label{eq:momexp}
\langle p^\alpha \rangle = \frac{2^{1-2\nu}Z^\alpha\sqrt{\pi}}{k!\eta^{\alpha}}
\frac{(k+\nu)\Gamma(k+2\nu)\Gamma(\nu+\frac{\alpha+1}{2})\Gamma(\nu+\frac{3-\alpha}{2})}
{\Gamma^2(\nu+\frac{1}{2})\Gamma(\nu+1)\Gamma(\nu+\frac{3}{z})}
\nonumber\\
\times \,_5F_4\left(
\left.
\begin{array}{ll}
-k,&k+2\nu,\nu,\nu+\frac{\alpha+1}{2},\nu+\frac{3-\alpha}{2}\\
&2\nu,\nu+\frac{1}{2},\nu+1,\nu+\frac{3}{2}
\end{array}
\right|
1
\right),
\end{eqnarray}
for the momentum expectation values \cite{Vanassche2000} of a generic $D$-dimensional hydrogenic state $(n,l)$. The latter expression is valid for $\alpha \in (-D-2l, D+2l+2)$, where $\nu\equiv L+1=l+\frac{D-1}{2}$ and $k=n-l-1$. The function ${}_{q}F_{p}(z)$ indicates the generalized hypergemetric series given by
\begin{equation}\label{eq:defhf}
	_{p+1}F_p\left(
\left.
\begin{array}{ll}
&a_1,...,a_{p+1}\\
&b_1,...,b_p
\end{array}
\right|
z
\right) = \sum_{j=0}^{\infty} \frac{(a_{1})_{j}\cdots(a_{p+1})_{j}}{(b_{1})_{j}\cdots(b_{p})_{j}}\frac{z^{j}}{j!}, 
\end{equation}
 which is terminating when the first one or more of the top parameters is a nonnegative integer, so that it reduces to a polynomial in $z$. Then, note that the previous functions ${}_3F_{2}(1)$ and ${}_5F_{4}(1)$ given by Eqs. (\ref{eq:posexp}) and (\ref{eq:momexp}), respectively, are terminating hypergeometric functions (so, polynomials) evaluated at $z=1$. Moreover, from Eq. (\ref{eq:posexp}) one obtains $\langle r^{0} \rangle = 1$ and 
\begin{eqnarray}
\label{eq:rvar}
\langle r\rangle &=& \frac{1}{2Z}[3\eta^{2}-L(L+1)], \quad
\langle  r^{2} \rangle = \frac{\eta^{2}}{2Z^{2}}[5\eta^{2}+1-3L(L+1)]\nonumber\\
\langle r^{-1} \rangle &=& \frac{Z}{\eta^{2}},\quad 
\langle r^{-2} \rangle = \frac{Z^{2}}{\eta^{3}}\frac{1}{L+\frac{1}{2}},\quad \langle r^{-3} \rangle = \frac{Z^{3}}{\eta^{3}} \frac{1}{L(L+\frac{1}{2})(L+1)} \nonumber \\
\langle r^{-4} \rangle &=& \frac{Z^{4}}{2\eta^{5}}\frac{3\eta^{2}-L(L+1)}{(L-\frac{1}{2})L(L+\frac{1}{2})(L+1)(L+\frac{3}{2})} \nonumber\\
\langle r^{-6} \rangle &=& \frac{Z^{6}}{8\eta^{7}} \frac{35\eta^2(\eta^2-1)- 30\eta^2(L+2)(L-1)+ 3(L+2)(L+1)L(L-1)}{(L-\frac{3}{2})(L-1)(L-\frac{1}{2})L(L+\frac{1}{2})(L+1)(L+\frac{3}{2})(L+2)(L+\frac{5}{2})}\nonumber\\ 
 \end{eqnarray}
for the first few position expectation values by means of the well-controlled properties of the function ${}_3F_{2}(1)$ \cite{Olver2010}. For Kramers-Pasternack two-term and three-term recurrence relations, see e.g. \cite{Ray1988} and Eq. (10.24) of \cite{Dong2011} for $D$-dimensional systems and the recent review \cite{Szymanski2020} for three-dimensional systems. For $D=3$ these multidimensional expressions reduce to the known three-dimensional ones \cite{Andrae1997,Drake1990,Hey1993b,Tarasov2004,Dehesa2010}.   \\

To calculate the corresponding quantities in momentum space from Eq. (\ref{eq:momexp}) is a formidable task, basically because ${}_5F_{4}(1)$ is a much more complicated hypergeometric function although it has a Saalschutzian or balanced character \cite{Slater1966,Lewanowicz1985}; that is, the sum of its lower parameters equals the sum of its upper parameters plus unity. For the particular cases $\alpha=0$ and $2$ the corresponding function ${}_5F_{4}(1)$ reduces to a ${}_3F_{2}(1)$-function which can be evaluated by means of the Pfaff-Saalschutz theorem, obtaining \cite{Slater1966,Vanassche2000}
\begin{equation}\label{eq:secondmom}
	\langle p^{0} \rangle = 1, \quad \langle p^{2}\rangle = \frac{Z^{2}}{\eta^{2}}.
\end{equation}
For values of $\alpha$ other than $0$ and $2$ we cannot use this approach because of the lack of appropriate reduction formula for the function ${}_5F_{4}(1)$. On the other hand, from Eq. (\ref{eq:moexpec}) one obtains in a straightforward manner the reflection or inversion formula \cite{Dehesa2010}
\begin{equation}\label{eq:reflectionform}
\left(\frac{\eta}{Z} \right)^{2-\alpha}\left\langle p^{2-\alpha} \right\rangle =\left(\frac{\eta}{Z}
\right)^{\alpha}\left\langle p^{\alpha} \right\rangle, \quad  \alpha = 0,1,2,... 
\end{equation}
which is not trivial for $\alpha \neq 1$. In particular, it gives Eq. (\ref{eq:secondmom}) and
\begin{equation}\label{eq:reflec}
\langle  p^{-1} \rangle	= \left(\frac{\eta}{Z}
\right)^{4} \langle  p^{3} \rangle, \quad \langle  p^{-2} \rangle	= \left(\frac{\eta}{Z}
\right)^{6} \langle  p^{4} \rangle,\quad \langle  p^{-3} \rangle	= \left(\frac{\eta}{Z}
\right)^{8} \langle  p^{5} \rangle
\end{equation}
for a few values of $\alpha$.

Then, a procedure based on some computer-algebra algorithms of Zeilberger type \cite{Koepf2014,Koepf2020} (see also \cite{Adkins} for the three-dimensional case) can also be used to obtain in a cumbersome way the momentum expectation values of even order higher than 2. Then, this procedure together with the reflection formula lead to the following expressions 
\begin{eqnarray}
\label{eq:pvar}
\langle p^{-2} \rangle &=& \frac{Z^{-2}}{\eta^{-2}}\frac{8\eta -3(2L+1)}{2L+1}, \quad 
\langle  p^{4} \rangle = \frac{Z^{4}}{\eta^{4}}\frac{8\eta -3(2L+1)}{2L+1},\nonumber\\
\langle  p^{6} \rangle &=& \frac{Z^{6}}{\eta^{6}}\frac{(4k+2\nu+1)(16k^2+40\nu k-4k+ 4 \nu^2 + 16 \nu+ 15)}{(2L+3)(2L+1)(2L-1)}.
\end{eqnarray}
Let us highlight, however, that the problem to calculate explicitly the momentum expectation values of odd integer order remain open for the $D$-dimensional hydrogenic system, except $\langle  p^{-1} \rangle$ for the three-dimensional hydrogen atom (i.e., $D=3, Z=1$)  \cite{Delbourgo2009} (see also \cite{Pathak1981} for a semiclassical estimation). They will be investigated later in Section \ref{sec:oddmomvalues}.\\

The approach described in this section does not work efficiently in the following two extreme cases: the highly-excited or Rydberg (i.e. $n\gg 1$) states and the high-dimensional or pseudo-classical  systems. The determination of the expectation values in these two extreme cases require a deeper knowledge of approximation theory and orthogonal polynomials \cite{Buyarov1999,Aptekarev2010}. Here below we collect the main results, recently found. 

\subsection{The high-energy (i.e. Rydberg) limit}

This is the quasi-classical limit in quantum physics since the wavelengths of the particles are small in comparison with the characteristic dimensions of the system and the wavefunctions of the quasi-classical state. To determine the position and momentum expectation values of arbitrary order which quantify the spreading of the Rydberg $D$-dimensional hydrogenic states (i.e., states where the electron has a high or very high principal quantum number $n$) is a formidable task. Indeed, in this case the highly oscillatory nature of the corresponding integrands (see Eqs. (\ref{eq:radexpec2}) and (\ref{eq:moexpec})) renders Gaussian quadrature ineffective as the number of quadrature points grows linearly with $n$ and evaluation of the involved high-degree polynomials are subject to round-off errors.  \\
The use of the weak-* asymptotics of Laguerre and Gegenbauer polynomials \cite{Buyarov1999} has allowed Aptekarev et al \cite{Aptekarev2010} to obtain the expressions  
\begin{equation}
\label{eq:rydr}
\langle r^{\alpha} \rangle = \left(\frac{\eta^{2}}{Z}\right)^{\alpha} \,\,\frac{2^{\alpha+1}\,\,\Gamma(\alpha+\frac{3}{2})}{\sqrt{\pi}\,\,\Gamma(\alpha+2)}\left(1+o(1)\right), \quad n\rightarrow \infty
\end{equation}
(with uniformly bounded $\alpha,l$ and $D$, and $\alpha >-3/2$) 
\begin{equation}
	\langle r^{-\alpha} \rangle \simeq \frac{Z^\alpha}{\eta^3}\,\frac{\Gamma(2L-\alpha+3)}{\Gamma(2L+\alpha)}\,\frac{2^{3\alpha-5}\Gamma(\alpha-3/2)}{\sqrt{\pi}\Gamma(\alpha-1)},\quad n\rightarrow \infty
\end{equation}
(valid for $3/2<\alpha<2L+3$) for the position expectation values, and  
\begin{eqnarray} \label{eq:p_alpha_n-infty2}
\langle p^{\alpha} \rangle &\simeq & \left( \frac{Z}{n} \right)^{\alpha}  \frac{\alpha-1}{ \sin \left(\frac{\pi}{2} (\alpha-1) \right)}\nonumber\\
 &=& \left( \frac{Z}{n} \right)^{\alpha} \frac{2}{\pi} \Gamma \left( \frac{\alpha+1}{2} \right)\Gamma \left( \frac{3-\alpha}{2} \right); \quad -1< \alpha < 3, \quad \alpha \neq 1,
\end{eqnarray}
\begin{equation} \label{eq:p_1_n-infty2bis}
 \langle p \rangle  \simeq \frac{2 Z}{\pi n}, \quad n\rightarrow \infty
\end{equation} 
for the momentum expectation values of the Rydberg hydrogenic states with $n\gg1$ and $l$ uniformly bounded. Note that Eqs. (\ref{eq:rydr}) and (\ref{eq:p_alpha_n-infty2}) provides, in particular, the exact values for the normalization $ \left( \left\langle r^0\right\rangle = \left\langle p^0\right\rangle = 1\right)$, the inverse radii $(\langle r^{-1} \rangle, \langle r^{-2} \rangle, \langle r^{-3} \rangle)$ and the kinetic energy $ \left( \left\langle p^2 \right\rangle = Z^2/n^2 \right)$ of the system, previously given; moreover, they give for instance that
\begin{equation}
\langle r^{-4} \rangle \simeq \frac{3Z^{4}}{2\eta^{3}(L-\frac{1}{2})L(L+\frac{1}{2})(L+1)(L+\frac{3}{2})},
\end{equation}
\begin{equation}
	\langle r \rangle \simeq \frac{3\eta^2}{2Z},\quad \langle r^2 \rangle \simeq \frac{5}{2} \left(\frac{\eta^2}{Z}\right)^2,
\end{equation}
which are in agreement with the corresponding asymptotics of the general Eqs. (\ref{eq:rvar}). However, the problem to find the ($n\rightarrow \infty$)-limit of the momentum expectation values $\left\langle p^{\alpha} \right\rangle$ with $\alpha \notin (-1,3)$ \textit{remains open} for the $D$-dimensional hydrogenic system, except $\left\langle p^{-1} \right\rangle$ which has been analyzed monographically for the three-dimensional hydrogen atom \cite{Delbourgo2009}. This problem will be discussed in section \ref{sec:oddmomvalues} for any multidimensional hydrogenic system.

Let us finally comment that the Rydberg position expectation values mentioned above include, extend to arbitrary  space dimensions and rigorously prove the corresponding semiclassical and quantum values obtained for three-dimensional systems \cite{Heim1994,Shiell2003}.

\subsection{The high-dimensional (i.e., pseudo-classical) limit} 

The pseudo-classical or high dimensional ($D\to\infty$) limit plays a relevant role in various scientific fields such as e.g. in physics of fluids \cite{Costigliola2019} and in quantum chemistry where it is the starting point of a very useful strategy, the D-dimensional scaling method of Herschbach et al \cite{Herschbach1993,Tsipis1996,Herschbach1996,Herschbach2020}, to study the atomic and molecular systems. This  method requires to solve a finite many-electron problem in the ($D\rightarrow\infty$)-limit and then, perturbation theory in $1/D$ is used to have an approximate result for the standard dimension ($D=3$), obtaining at times a quantitative accuracy comparable to the self-consistent Hartree-Fock calculations.\\
Here, we show the position and momentum expectation values for an arbitrary (but fixed) state $(n,l,\{\mu \})$ of $D$-dimensional hydrogenic systems when $D\rightarrow \infty$. We start from the general expressions (\ref{eq:posexp}) and (\ref{eq:momexp}), respectively. Then, the asymptotics of the ${}_3F_{2}(1)$ and ${}_5F_{4}(1)$ functions and various functions involving the gamma function has allowed Toranzo et al \cite{Toranzo2016} to claim 
\begin{eqnarray}\nonumber
\langle r^{\alpha} \rangle = \left(\frac{D^2}{4 Z}\right)^{\alpha }\left(1+\frac{(\alpha+1)(\alpha+4l-2)}{2D} \right)\left(1+\frac{(\alpha +1) (\alpha +2) (n-l-1)}{D+2 l-1}\right)\\\label{eq:averk1}
\times\quad  \left(1+\mathcal{O}\left(D^{-2} \right)\right)
\end{eqnarray}
(valid for $\alpha > -D-2l$) for the position expectation values, and
\begin{equation}
\label{eq:avepa}
\langle p^{\alpha} \rangle = \left(\frac{2Z}{D} \right)^{\alpha}\left(1+\frac{\alpha(\alpha-2)(2n-2l-1)}{2D} + \mathcal{O}(D^{-2})\right)
\end{equation}
(valid for $\alpha > -D-2l$) for the momentum expectation values of the $D$-dimensional hydrogenic system as $D\rightarrow \infty$, respectively. Note that in this limit one has that $\langle r^{\alpha} \rangle \rightarrow r_{char}^{\alpha}$ and $\langle p^{\alpha} \rangle \rightarrow p_{char}^{\alpha}$, where $r_{char} = \frac{D^2}{4 Z}$ and $p_{char} = \frac{D^2}{4 Z}$. Then, $r_{char}$ and $p_{char}$ appear to be as the characteristic length and momentum for this Coulomb problem. So, the high-dimensional hydrogenic system seems to behave like an electron moving with velocity $p_{char}$ in a circular orbit of angular momentum $\frac{D}{2}$ and radius $r_{char}$.

Finally, for the circular states $(l = n-1)$ the general expressions (\ref{eq:averk1})-(\ref{eq:avepa}) provide the following position and momentum expectation values
\begin{equation}
\label{eq:averacirc}
\langle r^{\alpha}\rangle(D,n,l=n-1) = \left(\frac{D^2}{4Z}\right)^{\alpha } \left[1+\frac{(\alpha +1) (4n+\alpha -6)}{2 D}\right]\left(1+\mathcal{O}\left(D^{-2} \right)\right)
\end{equation}
\begin{equation}
\label{eq:avepacirc}
\langle p^{\alpha} \rangle(D,n,l=n-1) = \left(\frac{2Z}{D} \right)^{\alpha}\left(1+\frac{\alpha(\alpha-2)}{2D} + \mathcal{O}(D^{-2})\right),
\end{equation}
respectively. Moreover, from these expressions we can easily obtain the position and momentum expectation values for the ground state ($n=1$) of the $D$-dimensional hydrogenic system at the $D\rightarrow \infty$ limit.

\section{The momentum expectation values: general expressions and applications}
\label{sec:oddmomvalues}

In the previous section we have shown how the momentum expectation values $\langle p^\alpha \rangle$ with an even integer $\alpha$ can be obtained in a closed form in terms of the hyperquantum numbers $(n,l)$, the dimensionality $D$ and the nuclear charge $Z$, by means of Eq. (\ref{eq:momexp}) and the reflection formula (\ref{eq:reflectionform})
\begin{equation}\label{eq:reflec1}
\langle  p^{-\beta} \rangle	= \left(\frac{\eta}{Z}\right)^{2\beta+2} \langle  p^{\beta+2} \rangle, \quad  \beta = 0,1,2,...
\end{equation}
which holds for $-2l-D-2 \le\beta\le 2l+D$.

Here we will find all the momentum expectation values $\langle p^\alpha \rangle$, $\alpha \in \mathbb{R}$, for any $D$-dimensional hydrogenic state $(n,l)$, making emphasis on the mostly unknown but interesting ones; namely, those with odd integer $\alpha$. Then, the resulting expressions are applied to the circular states. Later on, they are used to obtain the mean momentum    $\langle p \rangle$ and the average inverse momentum $\langle p^{-1} \rangle$ for general and some particularly relevant $D$-dimensional hydrogenic states.

We can do it either by using the general expressions given by Eqs. (\ref{eq:moexpec1})-(\ref{eq:moexpec}) or, alternatively and equivalently, by the expressions (\ref{eq:momexp}) given in integral or finite-sum ways, respectively. We start using Eq. (\ref{eq:momexp}) which by means of the definition (\ref{eq:defhf}) for the hypergeometric function ${}_5F_{4}(1)$ and the duplication formula of the gamma function (see e.g., \cite{Olver2010,Lewanowicz1985}), can be rewriten as
\begin{eqnarray}
\label{eq:moexpbis}
\langle p^{\alpha}\rangle &=& \left( \frac{Z}{\eta} \right)^{\alpha} \frac{2}{k!}\frac{(k+\nu)\Gamma(k+2\nu)}{\Gamma(2\nu+1)}\frac{\Gamma(\nu+\frac{\alpha+1}{2})\Gamma(\nu+\frac{3-\alpha}{2})}{\Gamma(\nu+\frac{1}{2})\Gamma(\nu+\frac{3}{2})}\nonumber \\
&\times & \sum_{j=0}^{k}(-1)^{j}{{k}\choose{j}}\frac{(k+2\nu)_{j}(\nu)_{j}(\nu+\frac{\alpha+1}{2})_{j}(\nu+\frac{3-\alpha}{2})_{j}}{(2\nu)_{j}(\nu+1)_{j}(\nu+\frac{1}{2})_{j}(\nu+\frac{3}{2})_{j}}
\end{eqnarray}
(valid for any real $\alpha \in (-D-2l, D+2l+2)$) for the momentum expectation values \cite{Vanassche2000} of a generic $D$-dimensional hydrogenic state $(n,l)$. Keep in mind that $\nu\equiv L+1=l+\frac{D-1}{2}$, $k=n-l-1$ and  the Pochhammer symbol $(a)_j = \Gamma(a+j)/\Gamma(a)$. This general expression includes, improves and generalizes to arbitrary dimensions the three-dimensional results previously obtained (see e.g., Eq. (13) of \cite{Hey1993}), which require a double summation. See also  \ref{alternative:app} where  it is shown a $D$-dimensional expression alternative and equivalent to (\ref{eq:moexpbis}), but suffering of a double summation.\\

To better manipulate the expression (\ref{eq:moexpbis}) it is convenient to take into account the following identities
\begin{eqnarray}
\frac{(2\nu+k)_{j}}{(2\nu)_{j}} =	\frac{(2\nu+j)_{k}}{(2\nu)_{k}}, \quad  \frac{(\nu)_{j}}{(\nu+1)_{j}} =	\frac{\nu}{\nu+j}
\end{eqnarray}
in Eq. (\ref{eq:moexpbis}), so that we can rewrite it as follows
\begin{eqnarray}
\label{eq:moexpbis2}
\langle p^{\alpha}\rangle  &=& \left( \frac{Z}{\eta} \right)^{\alpha}F_{k}(\nu,\alpha)\,f_{k}(\nu)
\end{eqnarray}
with 
\begin{equation}\label{eq:moexpbis3}
	F_{k}(\nu,\alpha) = \frac{2}{k!}\frac{(k+\nu)\Gamma(k+2\nu)}{\Gamma(2\nu+1)}\frac{\Gamma(\nu+\frac{\alpha+1}{2})\Gamma(\nu+\frac{3-\alpha}{2})}{\Gamma(\nu+\frac{1}{2})\Gamma(\nu+\frac{3}{2})}
\end{equation}
and
\begin{equation}
\label{eq:2}
f_{k}(\nu) = \frac{1}{(2\nu)_{k}}\sum_{j=0}^{k}(-1)^{j}{{k}\choose{j}}(2\nu+j)_{k}\,d_{j},
\end{equation}
where
\begin{eqnarray}
\label{eq:3}
d_{j}\equiv d_{j}(\nu) &=& \frac{\nu}{\nu+j}\frac{(\nu+\frac{\alpha+1}{2})_{j}(\nu+\frac{3-\alpha}{2})_{j}}{(\nu+\frac{1}{2})_{j}(\nu+\frac{3}{2})_{j}}.
\end{eqnarray}
Now, to illustrate the utility of this expression, let us calculate the momentum expectation values $\langle p^\alpha \rangle$ with any real $\alpha$ for the class of $D$-dimensional hydrogenic \textit{circular states $(n,l=n-1)$}. Here we have $\nu=\eta = n+ \frac{D-3}{2}$ and $k=0$, so that $d_0=1$, $f_{0}(\nu)=1$ and 
\begin{equation}\label{eq:F0eta}
	F_{0}(\nu,\alpha)= F_{0}(\eta,\alpha)
= \frac{\Gamma(\eta+\frac{\alpha+1}{2})\,\Gamma(\eta+\frac{3-\alpha}{2})}{\Gamma(\eta+\frac{1}{2})\,\Gamma(\eta+\frac{3}{2})}.
\end{equation}
Then, finally Eqs. (\ref{eq:moexpbis2})-(\ref{eq:moexpbis3}) and (\ref{eq:F0eta}) give
\begin{equation}\label{eq:moexpbis4}
\langle p^{\alpha}\rangle(D,n,l=n-1)  = \left( \frac{Z}{\eta} \right)^{\alpha}\frac{\Gamma(\eta+\frac{\alpha+1}{2})\,\Gamma(\eta+\frac{3-\alpha}{2})}{\Gamma(\eta+\frac{1}{2})\,\Gamma(\eta+\frac{3}{2})}
\end{equation}
for $\alpha \in (-D-2n+2, D+2n)	$. From this expression we can obtain, for example, the momentum expectation values $\langle p^{\alpha}\rangle(D,n=1,l=0)$ of the \textit{ground state} for any real $\alpha$ 
\begin{equation}
	\langle p^{\alpha}\rangle(D,n=1,l=0) = \left( \frac{2Z}{D-1} \right)^{\alpha}\frac{2\,\Gamma(\frac{D+\alpha}{2})\,\Gamma(\frac{D-\alpha+2}{2})}{D\,[\Gamma(\frac{D}{2})]^2}
\end{equation}
for $\alpha \in (-D, D+2)$. Moreover, we find the expressions
\begin{eqnarray}
\label{eq:moexpbis5}
\langle p^{\alpha}\rangle(D=3,n,l=n-1)  
&=& \left( \frac{Z}{n} \right)^{\alpha}\frac{\Gamma(n+\frac{\alpha+1}{2})\,\Gamma(n+\frac{3-\alpha}{2})}{\Gamma(n+\frac{1}{2})\,\Gamma(n+\frac{3}{2})}
\end{eqnarray}
(with $\alpha \in (-1-2n, -3+2n)$ for the momentum expectation values $\langle p^{\alpha}\rangle$ of the three-dimensional hydrogenic circular states, and 
\begin{eqnarray}
\left\langle p^\alpha \right\rangle(D=3,n=1,l=0)=\frac{8Z^\alpha  }{3 \pi}\,\Gamma\left(\frac{3+\alpha}{2} \right)\,\Gamma\left(\frac{5-\alpha}{2} \right); \alpha \in (-3, 5).\nonumber\\
\end{eqnarray}
for the corresponding quantities of the ground state of the three-dimensional hydrogenic atom. The last three-dimensional expression has been previously found by various authors (see e.g., pag. 325 of \cite{Thakkar2004}). Finally, from Eq. (\ref{eq:moexpbis5}) and the following asymptotics of the ratio of Gamma functions (see Eqs. 5.11.13 and 5.11.15 of \cite{Olver2010})
\begin{equation} \label{eq:asymp_gamma}
\frac{\Gamma(x+a)}{\Gamma(x+b)} \sim x^{a-b} \left[1 + \frac{1}{2x} (a-b)(a+b-1)\right]; \quad  x \rightarrow \infty,
\end{equation}
we obtain the expectation values
\begin{equation}\label{eq:alfaRyd}
\langle p^{\alpha}\rangle(D=3,n,l=n-1) \simeq \left(\frac{Z}{n}\right)^\alpha \left( 1+ \frac{\alpha(\alpha-2)}{4n} + O\left(\frac{1}{n^2}\right)\right); 
\, n\rightarrow \infty	
\end{equation}
for the Rydberg circular states of the three-dimensional hydrogenic states.

\subsection{The mean momentum $\langle p\rangle$}

Let us now apply this procedure based on Eq. (\ref{eq:moexpbis}) or Eq. (\ref{eq:moexpbis2}) to evaluate the mean momentum $\langle p \rangle$ for a generic $D$-dimensional hydrogenic state $(n,l)$, which is closely related to the Dirac exchange energy \cite{Daubechies1983}. For $\alpha = 1$ these general expressions give 
\begin{eqnarray}
\label{eq:1}
\langle p\rangle(D,n,l)  &=& \frac{Z}{\eta}\frac{2}{k!}\frac{(k+\nu)\Gamma(k+2\nu)}{\Gamma(2\nu+1)}\frac{\Gamma(\nu+1)\Gamma(\nu+1)}{\Gamma(\nu+\frac{1}{2})\Gamma(\nu+\frac{3}{2})}\nonumber \\
&\times & \sum_{j=0}^{k}(-1)^{j}{{k}\choose{j}}\frac{(k+2\nu)_{j}(\nu)_{j}(\nu+1)_{j}}{(2\nu)_{j}(\nu+\frac{1}{2})_{j}(\nu+\frac{3}{2})_{j}},
\end{eqnarray}
or equivalently,
\begin{eqnarray}
\label{eq:1bis}
\langle p\rangle(D,n,l) &=& \frac{Z}{\eta} \frac{2}{k!}\frac{(k+\nu)\Gamma(k+2\nu)}{\Gamma(2\nu+1)}\frac{\Gamma(\nu+1)\Gamma(\nu+1)}{\Gamma(\nu+\frac{1}{2})\Gamma(\nu+\frac{3}{2})}f_{k}(\nu)
\end{eqnarray}
 where $\nu\equiv L+1=l+\frac{D-1}{2}$, $k=n-l-1$, $\eta = n+ \frac{D-3}{2}$ and $f_{k}(\nu)$ is given by Eq. (\ref{eq:2}).\\

Operating as before we can calculate from these expressions the mean momentum value for  specific $D$-dimensional hydrogenic states. Then, we find
\begin{eqnarray}
	\left\langle p \right\rangle(D,n,l=n-1) &=& \frac{Z}{\eta} \frac{\Gamma(\eta+1)\,\Gamma(\eta+1)}{\Gamma(\eta+\frac{1}{2})\,\Gamma(\eta+\frac{3}{2})}\\
	 &=& \frac{4Z}{2n+D-3} \frac{[\Gamma(\frac{2n+D-1}{2})]^2}{(2n+D-2)\,[\Gamma(\frac{2n+D-2}{2})]^2}
\end{eqnarray}
for the circular states, and 
\begin{eqnarray}\label{eqI_cap1:<p><p2>_gs}
\left\langle p \right\rangle(D,n=1,l=0) &=& \frac{4Z}{D(D-1)}  \left[\frac{\Gamma \left( \frac{D+1}{2}\right) }{\Gamma \left( \frac{D}{2}\right)}\right]^2, 
\end{eqnarray}
for the ground state. Then, for the three-dimensional hydrogenic atom it follows that
\begin{eqnarray}\label{eq:meanmom5}
	\langle p\rangle(D=3,n,l=n-1) &=& \frac{Z}{n} \frac{\Gamma(n+1)\Gamma(n+1)}{\Gamma(n+\frac{1}{2})\Gamma(n+\frac{3}{2})}\\
	& =& \frac{Z}{\pi} \frac{2^{4n-1}n\,[(n-1)!]^4}{(2n+1)\,[(2n-1)!]^2}\nonumber
\end{eqnarray}
for the circular states, and  
\begin{equation}\label{eq:<pa>_gs_d3}
\left\langle p\right\rangle(D=3,n=1,l=0)=\frac{8Z}{3 \pi}
\end{equation}
for the ground state of the three-dimensional hydrogenic atom. Finally, from Eq. (\ref{eq:alfaRyd}) with $\alpha=1$ or from Eq. (\ref{eq:meanmom5}) and the asymptotics 
\begin{equation}
	\Gamma(az+b) \simeq (2\pi)^{1/2}\,e^{-az}\,(az)^{az+b-1/2};\quad a>0, \quad z\rightarrow \infty
\end{equation}
we find 
\begin{equation}
	\left\langle p \right\rangle (D=3,n,l=n-1) \simeq \frac{Z}{n}, \quad n\rightarrow \infty
\end{equation}
for the highly-excited (Rydberg) circular states of the three-dimensional hydrogenic states.\\

Alternatively, we can use Eq. (\ref{eq:moexpec1}) to obtain all the previous expressions and other ones as illustrated in  \ref{mean:app}. For instance, we obtain in addition the expression 
\begin{equation}
	\left\langle p \right\rangle (D=3,n,l=0)= \frac{2Z}{\pi n} \frac{4n^2}{4n^2-1}
\end{equation}
It is most interesting to remark that for the ground state $(n=1)$ one has the known exact value given by Eq. (\ref{eq:<pa>_gs_d3}), $\frac{8Z}{3\pi}$, and for the highly-excited (Rydberg) $(nS)$-wave states, which have $n\gg1$ and $l=0$, one has the expression (\ref{eq:p_1_n-infty2bis}), $\frac{2Z}{\pi n}$,
obtained by Aptekarev et al \cite{Aptekarev2010} by very different means, what is also a further checking of our results.

\subsection{The average inverse momentum $\langle p^{-1} \rangle$}
This quantity determines both the peak height of the Compton profile within the impulse approximation \cite{Pathak1986} and the energy level change induced by the Born reciprocity effect on all hydrogenic bound state levels which is important at atomic and galactic scales \cite{Delbourgo2009}. Here we apply the abovementioned procedure based on the general expressions (\ref{eq:moexpbis}) or Eq. (\ref{eq:moexpbis2}) to evaluate the average inverse momentum $\langle p^{-1} \rangle$ for a generic $D$-dimensional hydrogenic state $(n,l)$ and the expectation value $\langle p^3 \rangle$, which describes the  interelectronic repulsion energy as mentioned above. For $\alpha = -1$ these expressions give 
\begin{eqnarray}
\langle p^{-1}\rangle(D,n,l) &=&  \frac{\eta}{Z}  \frac{2}{k!}\frac{(k+\nu)\Gamma(k+2\nu)}{\Gamma(2\nu+1)}\frac{\Gamma(\nu)\Gamma(\nu+2)}{\Gamma(\nu+\frac{1}{2})\Gamma(\nu+\frac{3}{2})}\nonumber \\
&\times & \sum_{j=0}^{k}(-1)^{j}{{k}\choose{j}}\frac{(k+2\nu)_{j}(\nu)_{j}(\nu)_{j}(\nu+2)_{j}}{(2\nu)_{j}(\nu+1)_{j}(\nu+\frac{1}{2})_{j}(\nu+\frac{3}{2})_{j}}
\end{eqnarray}
or equivalently,
\begin{equation}\label{eq:inmom}
	\langle p^{-1}\rangle(D,n.l) =  \frac{\eta}{Z}  \frac{2}{k!}\frac{(k+\nu)\Gamma(k+2\nu)}{\Gamma(2\nu+1)}\frac{\Gamma(\nu)\Gamma(\nu+2)}{\Gamma(\nu+\frac{1}{2})\Gamma(\nu+\frac{3}{2})}\,f_{k}(\nu)
\end{equation}
 where $\nu\equiv L+1=l+\frac{D-1}{2}$, $k=n-l-1$, $\eta = n+ \frac{D-3}{2}$ and $f_{k}(\nu)$ is given by Eq. (\ref{eq:2}).\\
 From these expressions and operating as before, we can calculate the average inverse momentum value for  specific $D$-dimensional hydrogenic states. Then, we find the expressions
 \begin{eqnarray}\label{eq:circularminus1}
 	\langle p^{-1}\rangle(D,n,l=n-1) &=& \frac{\eta}{Z} \frac{\Gamma(\eta)\,\Gamma(\eta+2)}{\Gamma(\eta+\frac{1}{2})\,\Gamma(\eta+\frac{3}{2})}\\
 	& =& \frac{2n+D-3}{Z}\frac{\Gamma(\frac{2n+D-3}{2})\,\Gamma(\frac{2n+D+1}{2})}{(2n+D-2)\,[\Gamma(\frac{2n+D-2}{2})]^2}
 \end{eqnarray}
 for the circular states, and
 \begin{equation}
 	\langle p^{-1}\rangle(D,n=1,l=0) = \frac{D-1}{Z}\,\frac{\Gamma(\frac{D-1}{2})\,\Gamma(\frac{D+3}{2})}{D\,[\Gamma(\frac{D}{2})]^2}
 \end{equation}
 for the ground state. From the last two results we have the expressions 
\begin{eqnarray}
\label{eq:moexpbis6}
\langle p^{-1}\rangle(D=3,n,l=n-1) &=& \frac{n}{Z} \frac{\Gamma(n)\,\Gamma(n+2)}{\Gamma(n+\frac{1}{2})\,\Gamma(n+\frac{3}{2})}\\  
&=& Z^{-1}\, \frac{2\,\Gamma(n+1)\,\Gamma(n+2)}{(2n+1)\,[\Gamma(\frac{2n+1}{2})]^2},
\end{eqnarray}
\begin{equation} \label{eq:trimogs1}
\langle p^{-1}\rangle(D=3,n=1,l=0)  
= \frac{16}{3\pi Z} ,
\end{equation}
for the average inverse momentum of the circular states and the ground state of the hydrogenic atom, respectively. For the hydrogen atom (i.e., Z=1, D=3) the last two expressions are equal to the corresponding quantities obtained previously by Delbourgo and Elliott \cite{Delbourgo2009} by other means, what is a partial checking of our results.\\

 On the other hand, from Eq. (\ref{eq:inmom}) or better from Eq. (\ref{eq:moexpec1}) with $\alpha=-1$ (see \ref{inverse:app}),  we have also found the expression 
\begin{equation}\label{eq:trimogsn}
	\langle p^{-1}\rangle(D=3,n,l=0) = \frac{4n}{Z\pi} \left[\psi(n+\frac{1}{2})-\frac{2n^2}{4n^2-1}+\gamma+2\ln2    \right]
\end{equation}
for the average inverse momentum of the the three-dimensional hydrogenic ($nS$)-states, where $\psi(x)\equiv \Gamma'(x)/\Gamma(x)$ is the digamma function and $\gamma=0.5772...$ denotes the Euler-Mascheroni constant \cite{Olver2010}. Note that Eq. (\ref{eq:trimogsn}) reduces to Eq. (\ref{eq:trimogs1}) when $n=1$.   Moreover, taking into account Eq. (\ref{eq:asymp_gamma}) and the asymptotic behavior of the digamma $\psi(x)$ function \cite{Olver2010} 
\begin{equation}\label{eq:asymp_digamma}
\psi(x) \sim \log x -\frac{1}{2x}-\frac{1}{12x^2}; \qquad \qquad \textrm{for} \quad x \rightarrow \infty,
\end{equation}
we obtain from Eqs. (\ref{eq:circularminus1}) and (\ref{eq:trimogsn}) the following expectation values 
\begin{equation}
	\langle p^{-1}\rangle(D=3,n,l=n-1) \simeq \frac{n}{Z}\left(1+\frac{3}{4n}+ \mathcal{O}(n^{-2})\right); \quad n \rightarrow \infty
\end{equation}
 and 
$$\langle p^{-1}\rangle(D=3,n,l=0) \simeq \frac{4n}{\pi} \left[\log (4n)+ \gamma -\frac{1}{2}-\frac{1}{2n} -\frac{1}{12n^2} \right] + \mathcal{O}(n^{-3}); \quad n \rightarrow \infty,$$
for the Rydberg circular and $nS$ states of the hydrogen atom, respectively. See \cite{Delbourgo2009} for further news about this expectation value.\\
Alternatively, we can use Eq. (\ref{eq:moexpec1}) to obtain all the previous expressions and other ones as illustrated in  \ref{inverse:app}. Finally, let us highlight that by combining the previous expressions and the reflection formula (\ref{eq:reflec}) or (\ref{eq:reflec1}) one can obtain
\begin{equation}
	\langle  p^3 \rangle	= \left(\frac{Z}{\eta}\right)^4 \langle  p^{-1} \rangle
\end{equation}
for the corresponding multidimensional hydrogenic states in a straightforward manner. Working in this way, we can determine $\langle  p^{-3} \rangle$ and  $\langle  p^5 \rangle$, and the rest of momentum expectation values $\langle  p^{\alpha} \rangle$ with odd order $\alpha$.

\section{Position-momentum uncertainty properties}
\label{sec:posmomur}

In this section we show various mathematical formalizations of the position-momentum Heisenberg's uncertainty principle given by means of different inequality-based relations which connect the momentum radial expectation values with the position radial expectation values and the position entropic moments $W_{\alpha}[\rho]$ of the multidimensional quantum systems. These expectation values and entropic moments play a fundamental role  in the analysis of the structure and dynamics of natural systems and phenomena, from atomic and molecular systems to systems with non-standard dimensionalities \cite{Dong2011,Herschbach1993,Sen2012}.

The entropic moments of the position probability density are defined as
\begin{equation}
\label{eq:entropmom}
W_{\alpha}[\rho]=\int_{\mathbb R_D} [\rho(\vec{r})]^{\alpha}\,d^{D}\vec r,\quad \alpha\geq 1
\end{equation}
 These $\alpha\textit{th}$-quantities describe, save for a proportionality factor, numerous fundamental and experimentally accessible quantities \cite{Thakkar2004,Thakkar1990,Liu1996,Liu1997,Nagy1999}, such as e.g. the Dirac-Slater exchange energy $(\alpha = \frac{4}{3})$, Thomas-Fermi kinetic energy $(\alpha = \frac{5}{3})$,  the Patterson function of x-ray crystallography $(\alpha = 3)$, and others in the framework of the density functional theory \cite{Parr1989}.\\
 
 The position-momentum uncertainty relations are mathematical formulations of the quantum-mechanical uncertainty principle which describes a characteristic feature of quantum mechanics and states the limitations to perform measurements on a system without disturbing it. Indeed, the $D$-dimensional position-momentum uncertainty relation is the Heisenberg-like inequality \cite{Angulo1993,Angulo1994,Dehesa2010} 
 \begin{equation}
\langle r^a \rangle^{\frac{2}{a}} \langle p^b \rangle^{\frac{2}{b}} \, \ge \,
 \left( \frac{e \, D^{\frac{2}{a}} \,
\Gamma^{\frac{2}{D}}\left(1+\frac{D}{2} \right)}{(a e)^{\frac{2}{a}} \,
\Gamma^{\frac{2}{D}}\left(1+\frac{D}{a} \right)} \right) \, \left( \frac{e \,
D^{\frac{2}{b}} \, \Gamma^{\frac{2}{D}}\left(1+\frac{D}{2} \right)}{(b
e)^{\frac{2}{b}} \, \Gamma^{\frac{2}{D}}\left(1+\frac{D}{b} \right)} \right),
\label{eq:rAlfapBeta}
\end{equation}
valid for  all $(a,b)\in\ \mathbb{R}_2^+=(0,+\infty)^2$. This relation, obtained by using information-theoretic methods, was later improved by Zozor et al \cite{Zozor2011}. See also \cite{Folland1997} and \cite{Angulo2011} for similar inequalities involving modified expectation values and expectation values with negative orders, respectively.
For $a=b=2$, this expression boils down to the familiar $D$-dimensional form of the inequality
\begin{equation}\label{eq:r2p2}
\left\langle r^2\right\rangle \left\langle p^2\right\rangle \geq \frac{D^2}{4},
\end{equation}
firstly found by Heisenberg and Kennard for one-dimensional systems \cite{Heisenberg1927,Kennard1927}. Later on, this expression has been improved for the spherically-symmetric potentials \cite{Sanchez2006} as
\begin{equation}\label{eq:r2p2sphe}
\left\langle r^2\right\rangle \left\langle p^2\right\rangle \geq \left(l + \frac{D}{2}\right)^2,
\end{equation}
where $l$ denotes the orbital hyperangular quantum number given in Section \ref{sec:probdensities}. \\

For completeness let us quote here that Eq. (\ref{eq:rAlfapBeta}) for $D=3$ can be cast in the form
\begin{equation*}
\left\langle r^{a}\right\rangle^{1/a} \left\langle p^{b}\right\rangle^{1/b} \geq \left( \frac{\pi a b}
{16 \Gamma \left(\frac{3}{a} \right)\Gamma \left(\frac{3}{b} \right)} \right)^{1/3} \left( \frac{3}{a}\right)
^{\frac{1}{a}}\left( \frac{3}{b}\right)^{\frac{1}{b}} e^{1-\frac{1}{a}-\frac{1}{b}}; \hspace{2mm} a>0,b>0
\end{equation*}
which for the specially interesting case $a=b>0$ takes the form
\begin{equation}\label{eq:rapa}
\left\langle r^{a}\right\rangle \left\langle p^{a}\right\rangle \geq \left\lbrace \left( \frac{27 \pi}{16 a \Gamma \left( \frac{3}{a} \right)} \right)^{\frac{1}{3}}
 \left( \frac{a e}{3} \right)^{1-\frac{2}{a}} \right\rbrace^a, \hspace{2cm} a>0.
\end{equation}
Note that for $a=2$ this inequality reduces to Eq. (\ref{eq:r2p2}) with $D=3$.

In addition, from the Pitt-Beckner inequality it is possible to find \cite{Beckner1995,Dehesa2007njp} the following relation 
between the expectation values $\left<p^\alpha\right>$ and $\left< r^{-\alpha}\right>$ of 
$D$-dimensional systems:
\begin{equation}\label{eq:pAlfa}
\left<p^\alpha\right> \ge 2^\alpha \left[
\frac{\Gamma\left( D+\frac{\alpha}{4} \right)}
{\Gamma\left(D-\frac{\alpha}{4}\right)}
\right]^2 \left< r^{-\alpha}\right>; \qquad 0\le\alpha<D
\end{equation}
which for $\alpha=2$ allows us to obtain another bound for the kinetic energy; namely, 
\begin{equation}
T\ge \frac{(D-2)^2}{8}\left< r^{-2}\right>; \qquad D>2.
\end{equation}
It is straightforward, although at times a bit cumbersome, to show that the position and momentum expectation values previously obtained for the $D$-dimensional hydrogenic system fulfill the Heisenberg-like uncertainty relations here derived.

Recently, it has been argued that the  momentum expectation values and the position entropic moments for $D$-dimensional systems of $N$ fermions with spin $s$ fulfill the following semiclassical spin-dependent uncertainty-like relations of Daubechies-Thakkar type \cite{Daubechies1983,Thakkar2004,Thakkar1990} (see also \cite{Toranzo2015}):
 \begin{equation}
 \label{eq:avevalmom4}
 \langle p^{k}\rangle \geq K_{D}(k)q^{-\frac{k}{d}} W_{1+\frac{k}{D}}[\rho],
 \end{equation}
 where $k>0$, $q = 2s + 1$ is the number of spin states, and
 \begin{equation}
 \label{eq:Kdk}
  K_{D}(k) = \frac{D}{k+D}(2\pi)^{k}\frac{\left[\Gamma\left(1+\frac{D}{2}\right)\right]^{k/D}}{\pi^{k/2}}\, .
 \end{equation}
 One can realize that for $k<0$, the sign of inequality (\ref{eq:avevalmom4}) is inverted. Note that these expressions simplify for three-dimensional systems as
 \begin{equation}
\label{eq:empiricavevalmom2}
\langle p^{k}\rangle \leq c_{k} W_{1+\frac{k}{3}}[\rho] \quad \textrm{for} \quad k=-2,-1
\end{equation}
and
\begin{equation}
\label{eq:empiricavevalmom3}
\langle p^{k}\rangle \geq c_{k} W_{1+\frac{k}{3}}[\rho] \quad \textrm{for} \quad k=1,2,3,4
\end{equation}
with $c_{k}=3(3\pi^{2})^{k/3}(k+3)^{-1}$ (since $K_{d}(k) = 2^{\frac{k}{3}}c_{k}$ for $D =3$ and $s=1/2$), which were previously found by means of numerous semiclassical and Hartree-Fock-like ground-state calculations in atoms and diatomic molecules \cite{Pathak1986, Thakkar2004, Hart2005, Thakkar1990, Porras1990, Porras1995}. Moreover, the case $k = 2$ in Eq. (\ref{eq:avevalmom4}) was previously conjectured by Lieb (see e.g. \cite{Lieb1983}), and weaker versions of it have been rigorously proved as discussed elsewhere \cite{Thakkar1990}. In fact, Eq. (\ref{eq:avevalmom4}) with constant $K'_{D}(k) = K_{D}(k) \times B(D,k)$ with $B(D,k) = \left\{\Gamma\left(\frac{D}{k}\right)\inf_{a>0}\left[ a^{-\frac{D}{k}}\left(\int_{a}^{\infty}du\, e^{-u}(u-a)u^{-1}\right)^{-1}\right] \right\}^{-\frac{k}{D}}$ has been rigorously proved by Daubechies \cite{Daubechies1983}. Let us also mention that a number of authors have published some rigorous $D$-dimensional bounds of the same type \cite{Lieb2010, Hundertmark2007} with much less accuracy.\\

Furthermore, the use of the lower bound (\ref{eq:avevalmom4}) for the expected value of $\langle p^k \rangle$ together with the variational bounds to the involved entropic moments \cite{Dehesa1988,Dehesa1989,Galvez1987} has led to find \cite{Toranzo2014,Toranzo2015} spin-dependent Heisenberg-like uncertainty relations for $N$-fermion systems of the following type
\begin{equation}
\label{eq:genheisuncprod}
\langle r^{\alpha}\rangle^{\frac{k}{\alpha}}\langle p^{k}\rangle \geq \mathcal{F}(D,\alpha,k)\,q^{-\frac{k}{D}}N^{1+k \left(\frac{1}{\alpha }+\frac{1}{D}\right)},
\end{equation}
with $\mathcal{F}(D,\alpha,k) = K_{D}(k)F(D,\alpha,k)$ and
\begin{eqnarray}
\label{eq:F}
F(d,\alpha,k) & = & \frac{\left(1+\frac{k}{D}\right)^{1+\frac{k}{D}}\alpha^{1+\frac{2k}{D}}}{\left[\Omega_{D}B\left(\frac{D}{\alpha},2+\frac{D}{k}\right)\right]^{\frac{k}{D}}}\times  \left\{\frac{k^{k}}{\left[\left(1+\frac{k}{D}\right)\alpha+k\right]^{\left(1+\frac{k}{D}\right)\alpha+k}}\right\}^{\frac{1}{\alpha}},
\end{eqnarray}
where $\Omega_{D}=\frac{2\pi^{D/2}}{\Gamma(D/2)}$ is the volume of the unit hypersphere. This uncertainty relation, which holds for all  systems of $N$ fermions with spatial dimensionality $d$ and spin dimensionality $q=2s+1$, was previously found for $k=2$ by means of the Lieb-Thirring inequality \cite{Toranzo2014}. Moreover, for $D=3$ and $q=2$ it gives
\begin{equation}
\label{eq:genheisuncprod2}
\langle r^{\alpha}\rangle^{\frac{k}{\alpha}}\langle p^{k}\rangle \geq \mathcal{F}(3,\alpha,k)\,2^{-\frac{k}{3}}N^{\frac{k}{\alpha }+\frac{k+3}{3}},
\end{equation}
which holds for all N-electron systems. From this electronic Heisenberg's uncertainty relation many instances can be found, such as e.g. for $\alpha=k=2$ one has $\langle r^{2}\rangle\langle p^{2}\rangle \geq 1.85733 \times q^{-\frac{2}{3}}N^{\frac{8}{3}} = 1.17005 N^{\frac{8}{3}}$. Moreover, uncertainty relations similar to Eq. (\ref{eq:genheisuncprod}) have been recently found \cite{Toranzo2016jpa} by means of the extremization of various information-theoretic measures.

Finally, for completeness, let us also point out that similar uncertainty relations based on position and momentum entropic moments have been also found \cite{Angulo2012jmp,Guerrero2011} by means of the R\'enyi-entropy-based uncertainty inequalities \cite{Bialynicki2006,Zozor2008}.

\section{Conclusions}
The theoretical analysis of numerous fundamental and experimentally accessible quantities of multidimensional quantum systems leans heavily on accurate knowledge of the position and momentum radial expectation values.
In this work we have analytically studied and partially reviewed the radial expectation values of the multidimensional hydrogenic system in the two conjugated position and momentum spaces. These quantities have been expressed for all discrete stationary $D$-dimensional hydrogenic states in a closed and compact manner by means of  the dimensionality, the nuclear charge and the states' hyperquantum numbers.\\

Emphasis has been placed on the most unknown but interesting ones: the momentum radial expectation values. The usefulness of the resulting general expressions is illustrated by their application to some classes of multidimensional hydrogenic states of circular and ($nS$)-wave character, as well as to the high-energy (Rydberg) and high-dimensional (pseudoclassical) states which are relevant \textit{per se} and because they play an important role in numerous phenomena of the multidimensional quantum physics. Finally, we have given various not-so-well-known mathematical formalizations of the position-momentum Heisenberg's uncertainty principle of the multidimensional quantum physics which are expressed in terms of the position an momentum expectation values and the entropic moments of the state's probability density. They are fulfilled by the radial expectation values of the multidimensional hydrogenic system considered in this work.

%

\section*{Acknowledgments}
This work has been partially supported by the Grant FIS2017-89349P of the Agencia Estatal de Investigaci\'on (Spain) and the European Regional Development Fund (FEDER).


\appendix

\section{Alternative $D$-dimensional expression for the momentum expectation values}
\label{alternative:app}

Here we derive an alternative expression for the general momentum expectation values $\langle p^{\alpha} \rangle$ of the $D$-dimensional hydrogenic states. This expression is equivalent to (\ref{eq:moexpbis}), but it is suffering of a double summation. We begin with Eqs. (\ref{eq:moexpec1}) and (\ref{eq:karara}), obtaining
\begin{eqnarray}
\langle p^{\alpha} \rangle &=& \frac{K_{n,l}'^{2}}{2^{2L+4}}\left(\frac{Z}{\eta}\right)^{\alpha+D}\left(\frac{\Gamma(L+\frac32)}{\Gamma(2L+2)}\frac{\Gamma(n+l+D-2)}{\Gamma(n+\frac D2-1)}\right)^2 \nonumber \\
&\times & \int_{-1}^{1} (1-y)^{l+\frac{\alpha+D}{2}-1} (1+y)^{l+\frac{D-\alpha}{2}}  \left[\mathcal{P}_{n-l-1}^{(L+\frac12,L+\frac12)}\left(y\right)\right]^{2}dy, 
\end{eqnarray} 
where we have taken into account the following relation of the Gegenbauer and Jacobi polynomials \cite{Olver2010}
\begin{equation}
	\mathcal{C}_{n}^{(\lambda)}(x) = \frac{(2\lambda)_n}{(\lambda+\frac{1}{2} )_n}\mathcal{P}_{n}^{(\lambda-\frac12,\lambda-\frac12)}\left(x\right).
\end{equation}
Now we use the known formula for the Jacobi polynomials \cite{Srivastava1988,Sanchez-Moreno2013} 
\begin{eqnarray*}
\hspace{-2cm}\int_{-1}^1 (1-y)^a(1+y)^b [\mathcal{P}_n^{(\gamma,\gamma)}(y)]^2\,dy &=&\sum_{i=0}^\infty \tilde c_i(0,2,n,\gamma,\gamma,a,b)\times
\\
&\times&\int_{-1}^1 (1-y)^a(1+y)^b\,\mathcal{P}_i^{(a,b)}(y)\,dy
\\
&=&\tilde c_0(0,2,n,\gamma,\gamma,a,b)\,\frac{2^{a+b+1}}{a+b+1}\,\frac{\Gamma(a+1)\Gamma(b+1)}{\Gamma(a+b+1)}
\end{eqnarray*}
with $a,b>-1$, and
\begin{equation*}
\hspace{-2cm}
\tilde c_{0}\left(0,2,n,a,a,\gamma,\delta\right)= {{n+a}\choose{n}}^{2} \sum_{j_1,j_2=0}^{n}\frac{(-n)_{j_1}(n+2a)_{j_1}(-n)_{j_2}(n+2a)_{j_2}}{(a+1)_{j_1}(a+1)_{j_2} j_1! j_2 !}\,\frac{(\gamma+1)_{j_1+j_2}}{(\gamma+\delta+2)_{j_1+j_2}}\\
\end{equation*}
Then, the expression reduces as follows
\begin{equation*}
\hspace{-2cm}\langle p^\alpha \rangle = 4\eta\left(\frac{Z}{\eta}\right)^{\alpha}\frac{ \Gamma(l+\frac{D-\alpha}{2}+1)}{\,\Gamma(n+l+D-2)\,\Gamma(n-l)}	\sum_{i,j=0}^{n-l-1}\Pi_{i,j}(n,l,D)\,\Gamma\left(l+\frac{D+\alpha}{2}+i+j\right),
\end{equation*}
where
%
\begin{equation*}
\hspace{-2cm}\Pi_{i,j}(n,l,D)=\frac{(-n-l+1)_{i}(-n-l+1)_{j}\Gamma(n+l+D-2+i)\Gamma(n+l+D-2+j)}{\Gamma(l+\frac D2+i)\Gamma(l+\frac D2+j)\Gamma(2l+D+1+i+j)\,i!\,j! }
\end{equation*}
which holds for $\alpha\in\left(-D-2l,\,D+2l+2\right)$.

\section{Mean momentum of multidimensional hydrogenic states} 
\label{mean:app}
Alternatively, to calculate $\langle p \rangle$ for a generic $D$-dimensional hydrogenic state we can use the general expressions given by Eq. (\ref{eq:moexpec1})-(\ref{eq:moexpec}). We first put $\alpha=1$ in the previous expressions, obtaining
\begin{eqnarray}
\left\langle p \right\rangle(D,n,l)
&=& \mathcal{K'}_{n,l}\int_{-1}^{+1}  \left[ {\cal{C}}_{\eta-L-1}^{(\nu)}(t) \right]^2 \left(  1-t^2 \right)^{\nu} dt 
\label{eq:meanmo1}\\
&=& \frac{Z}{\eta} \int_{-1}^{+1} \omega_{\nu}^{*}(t) \left[ {\tilde{\cal{C}}}_{\eta-L-1}^{(\nu)}(t) \right]^2 \left(  1-t^2 \right)^{\frac{1}{2}}dt\\
&=& \frac{Z}{\eta} \int_{-1}^{+1}  \left[ {\tilde{\cal{C}}}_{\eta-L-1}^{(\nu)}(t) \right]^2 \left(  1-t^2 \right)^{\nu}dt
\end{eqnarray}
with $\nu = L+1=l + \frac{D-1}{2}$, $\eta = n+ \frac{D-3}{2}$, $k=\eta- L -1 =n-l-1$, and 
\begin{eqnarray}\label{eq:mo1cte}
	\mathcal{K'}_{n,l} &=& \frac{Z}{\eta} \mathcal{K}_{n,l} = \frac{Z K_{n,l}'^{2}}{2^{2l+D+1} \eta^{D+1}} = Z\, 2^{2(L+1)}\left[\Gamma(L+1)\right]^2 \left(\frac{(\eta-L-1)!}{2\pi(\eta+L)!}\right)\nonumber \\
	&=& Z\, 2^{2l+D-1}\left[\Gamma\left(l+\frac{D-1}{2}\right)\right]^2 \left(\frac{(n-l-1)!}{2\pi(n+l+D-3)!}\right)
\end{eqnarray}  
Then, from Eq. (\ref{eq:meanmo1}) one has that the mean momentum of an arbitrary state $(n,l)$ of the $D$-dimensional hydrogenic system is given by
\begin{equation} \label{eq:meanmo2bis}
\left\langle p \right\rangle(D,n,l) =  \mathcal{K'}_{n,l}\,\,\mathcal{I}_{n,l}
\end{equation}
with the integral 
\begin{equation}
\mathcal{I}_{n,l} = \int_{-1}^{+1}  \left[ {\cal{C}}_{n-l-1}^{(\nu)}(t) \right]^2 \left(  1-t^2 \right)^{\nu} dt	
\end{equation}
with $\nu = L+1=l + \frac{D-1}{2}$. This integral can be expressed as a double-sum expression with the procedure described in the previous appendix, or better by means of the single-sum  expression (\ref{eq:1bis}) mentioned above. Here below, for illustration, we show the value of the mean momentum in these two cases for the three-dimensional (i.e., $D=3$) hydrogenic atom.

\subsection{$(nS)$ states}

In this case $D=3$ and $l=0$ so that $\nu=1$. Then, the involved integral $\mathcal{I}_{n,l}$ has the value 
\begin{equation}\label{eq:in0}
	\mathcal{I}_{n,0} = \frac{1}{2^{2n}(n!)^2}  \int_{-1}^{+1}  \left[ \frac{d^n}{dt^n}  \left(  1-t^2 \right)^{-\frac{1}{2}}\right]^2 dt =\frac{4n^2}{4n^2-1},
\end{equation}
we have, from Eqs. (\ref{eq:mo1cte})-(\ref{eq:in0}), that
\begin{equation}
	\left\langle p \right\rangle (D=3,n,l=0)=  \mathcal{K'}_{n,0}\,\,\mathcal{I}_{n,0} = \frac{2Z}{\pi n} \frac{4n^2}{4n^2-1}
\end{equation}
It is most interesting to remark that for the ground state $(n=1)$ one has the known exact value \cite{Dehesa2010} $\frac{8Z}{3\pi}$, and for the highly-excited (Rydberg) $S$-wave states, which have $n\gg1$, one has the expression (\ref{eq:p_1_n-infty2bis}), $\frac{2Z}{\pi n}$,
obtained by Aptekarev et al \cite{Aptekarev2010} by very different means, what is also a further checking of our results.

\subsection{Circular states}
In this case $l=n-1$ so that $k=0$ and $\nu = l+1= n$. Then, the integral $\mathcal{I}_{n,n-1}$ is
\begin{equation}
\mathcal{I}_{n,n-1} = \int_{-1}^{+1}  \left(  1-t^2 \right)^{n} dt = 2\frac{(2n)!!}{(2n+1)!!} = 2^{2n+1}\frac{(n!)^2}{(2n+1)!}
\end{equation}
since the involved Gegenbauer polynomial collapse to unity. Here, the symbols $(2n)!! = 2\times 4\times....\times 2n$ and $(2n-1)!! = 1\times3\times .....\times (2n-1)$, and the constant
\begin{equation}
	 \mathcal{K'}_{n,n-1} = Z\, 2^{2n-1}\frac{(n-1)!^2}{\pi(2n-1)!}.
\end{equation}
Finally, from Eqs. (\ref{eq:mo1cte})-(\ref{eq:meanmo2bis}) we have
\begin{eqnarray} \label{eq:meanmo2}
\left\langle p \right\rangle (D=3,n,l=n-1) &=&  \mathcal{K'}_{n,n-1}\,\,\mathcal{I}_{n,n-1}\\
 &=& Z\, 2^{4n-1}\frac{n\,[(n-1)!]^4}{\pi(2n+1)[(2n-1)!]^2},
\end{eqnarray}
which equals to the previous Eq. (\ref{eq:meanmom5}). 

\section{Average inverse momentum of multidimensional hydrogenic states} 
\label{inverse:app}
Alternatively, we can also calculate $\langle p^{-1} \rangle$ for a generic $D$-dimensional hydrogenic state by means of the general expressions given by Eq. (\ref{eq:moexpec1})-(\ref{eq:moexpec}). We first put $\alpha=-1$ in the previous expressions, obtaining
\begin{eqnarray}
\left\langle p^{-1} \right\rangle (D,n,l)
&=& \mathcal{K}_{n,l}''(D)\int_{-1}^{+1}  \left[ {\cal{C}}_{k}^{(\nu)}(t) \right]^2 \left(  1-t \right)^{\nu-1} \left(  1+t \right)^{\nu+1}dt
\label{eq:moexpec6}\\
&=& \frac{\eta}{Z} \int_{-1}^{+1} \omega_{\nu}^{*}(t) \left[ {\tilde{\cal{C}}}_{k}^{(\nu)}(t) \right]^2 \left(  1-t \right)^{-\frac{1}{2}} \left(1+t \right)^{\frac{3}{2}}dt.
\label{eq:moexpec7}
\end{eqnarray}
Note that $k = \eta - L -1 = n-l-1$, $\nu = L+1 = l + (D-1)/2$, $\omega_\nu^* (t)=(1-t^2)^{\nu-\frac{1}{2}}=(1-t^2)^{l+\frac{D-2}{2}}$ is the weight function of the Gegenbauer polynomials ${\tilde{\cal{C}}}_{k}^{(\nu')} (t)$, and the constant
\begin{eqnarray}
	\hspace{-2cm}\mathcal{K}_{n,l}''(D) &=&\frac{\eta}{Z}\mathcal{K}_{n,l} =\frac{\eta^2}{Z} \, 2^{2(L+1)}\left[\Gamma(L+1)\right]^2 \left(\frac{(\eta-L-1)!}{2\pi(\eta+L)!}\right)\nonumber\\ 
	&=& \frac{(2n+D-3)^2}{Z} \, 2^{2l+D-3}\left[\Gamma\left(l+\frac{D-1}{2}\right)\right]^2 \frac{(n-l-1)!}{2\pi(n+l+D-3)!}.
\end{eqnarray}
Then, the average inverse momentum of an arbitrary state $(n,l)$ of the $D$-dimensional hydrogenic system is given by
\begin{equation} \label{eq:meanmo7}
\left\langle p^{-1} \right\rangle(D,n,l) =  \mathcal{K}_{n,l}''(D)\,\,\mathcal{J}_{n,l}(D)
\end{equation}
with the integral 
\begin{equation}
\mathcal{J}_{n,l}(D) = \int_{-1}^{+1}  \left[ {\cal{C}}_{n-l-1}^{(\nu)}(t) \right]^2 \left(  1-t^2 \right)^{\nu-1}\left(  1+t \right)^{2} dt	
\end{equation}
with $\nu = L+1=l + \frac{D-1}{2}$. The explicit solution of this value was given above by means of Eq. (\ref{eq:moexpbis}). This integral can be expressed as a double-sum expression with the procedure described in the \ref{alternative:app}, or better by means of the single-sum  expression (\ref{eq:inmom}) mentioned above.

For the three-dimensional hydrogenic atom the last two expressions lead to
\begin{eqnarray} \label{eq:meanmo8}
\left\langle p^{-1} \right\rangle(D=3,n,l) &=&  \mathcal{K}_{n,l}''(D=3)\,\,\mathcal{J}_{n,l}(D=3)\nonumber\\
&=& \frac{2n^2}{Z\pi} \, (2^{l} l!)^2 \frac{(n-l-1)!}{(n+l)!}\int_{-1}^{+1}  \left[ {\cal{C}}_{n-l-1}^{(l+1)}(t) \right]^2 \left(1-t^2 \right)^{l}\left(  1+t \right)^{2} dt\nonumber\\	
\end{eqnarray}
which was previously found \cite{Delbourgo2009}, what is a further checking of our results. For illustration, the application of this expression to the {$(nS)$ states allows us to obtain 
\begin{eqnarray}
\left\langle p^{-1} \right\rangle(D=3,n,l=0) &=& \frac{2n}{Z\pi} \int_{-1}^{+1}  \left[ {\cal{C}}_{n-1}^{(1)}(t) \right]^2 \left(  1+t \right)^{2} dt\nonumber\\
&=&	\frac{4n}{Z\pi} \left[\psi(n+\frac{1}{2})-\frac{2n^2}{4n^2-1}+\gamma+2\ln2    \right]
\end{eqnarray}
as mentioned above, since $l=0$ so that $\nu=1$ in this case. See \cite{Delbourgo2009} for further details about
the average inverse momentum of the hydrogen atom.


\section*{References}

\begin{thebibliography}{88}



\bibitem{Dong2011} S. H. Dong,  \textit{Wave Equations in Higher Dimensions}, Springer, Berlin, 2011.
\bibitem{Yanez1994} R. J. Y\'a\~nez, W. Van Assche, J. S. Dehesa, Phys. Rev. A \textbf{50} (1994) 3065.
\bibitem{Aquilanti1997} V. Aquilanti, S. Cavalli, C. Coletti, Chem. Phys. \textbf{214} (1997) 1.
\bibitem{Avery2006} J. Avery, J. Avery, \textit{Generalized Sturmians and Atomic Spectra}, World Sci. Publ., New York, 2006.
\bibitem{Bonham1974} R. A. Bonham, M. Fink, High Energy Electron Scattering, Van Nostrand Reinhold: New York, 1974.
\bibitem{Williams1977} B. G. Williams (Ed), Compton Scattering: The Investigation of Electron Momentum Distributions, McGraw-Hill: New York, 1977.
\bibitem{Thakkar1980} A. J. Thakkar, A. M. Simas, V. H. Smith Jr., Extraction of momentum expectation values from Compton profiles, Mol. Phys. \textbf{41} (1980) 1153.
\bibitem{Glusker1986} J. P. Glusker, B. K. Patterson, M. Rossi (Eds), Patterson and Pattersons, Oxford University Press: New York, 1986.
\bibitem{Coppens1997} P. Coppens, X-ray Charge Densities and Chemical Bonding, Oxford University Press: New York, 1997.
\bibitem{Weigold1999} E. Weigold, I. E. McCarthy, Electron Momentum Spectroscopy, Kluwer Academic: New York, 1999.

\bibitem{Thakkar2004} A. J. Thakkar, \textit{Adv. Chem. Phys.} {\bf 128} (2004) 303.

\bibitem{Thakkar2005} A. J. Thakkar, Electronic structure: the momentum perspective. in: C.E. Dykstra, G. Frenking, K. S. Kim, G.E. Scuseria (Eds), \textit{Theory and Applications of Computational Chemistry: The First 40 Years}, Amsterdam, North Holland, 2005, chapter 19, , p. 483. 
\bibitem{Thakkar1990} A. J. Thakkar, W.A. Pedersen, \textit{Int. J. Quant. Chem.: Quant. Chem. Symp.} {\bf 24}, 327 (1990).
\bibitem{Hart2005} J. R, Hart, A.J. Thakkar, \textit{Int. J. Quantum Chem.} {\bf 102} (2005) 673.
\bibitem{Blair2014} S. A. Blair, A.J. Thakkar, \textit{J. Chem. Phys.} {\bf 141} (2014) 074306.
\bibitem{Gadre1991} S. R. Gadre, R. K. Pathak, Adv. Quantum Chem., \textbf{22} (1991) 211.
\bibitem{March1987} N. H. March, B.M. Deb. \textit{The Single-Particle Density in Physics and Chemistry}, Academic Press, New York, 1987. 

\bibitem{Lehtola2013} S. Lehtola, \textit{Computational Modelling of the Electron Momentum Density}, University of Helsinki: Ph. D. Thesis, 2013.

%


\bibitem{Delbourgo2009} R. Delbourgo, D. Elliott, J. Math. Phys. \textbf{50} (2009) 062107.

\bibitem{Aquilanti2001} V. Aquilanti, S. Cavalli, C. Coletti, D. D. Domenico, G. Grossi, Int. Rev. Phys. Chem. 20 (2001) 673.
\bibitem{Dehesa2011} J. S. Dehesa, S. L{\'{o}}pez-Rosa, D. Manzano, Entropy and complexity analysis of $D$-dimensional quantum systems, in: K.D. Sen (Ed), \textit{Statistical Complexities: Application to Electronic Structure}, Springer, Berlin, 2012. Chapter 5.
 \bibitem{Dehesa2010} J.S. Dehesa, S. L\'opez-Rosa,  A. Mart\'inez-Finkelshtein, R. J. Y\'a\~nez, \emph{Int. J. Quant. Chem.} {\bf 110} (2010) 1529. 
  \bibitem{Coletti2013} C. Coletti, D. Calderini, V. Aquilanti, Adv. Quantum Chem. \textbf{67} (2013) 73.
  \bibitem{Dulieu2018} O. Dulieu, J. Colgan, E. Grant, E. Krishnakumar, A. Osterwalder, H. Sadeghpour, M. Vrakking, J. Wu (Eds), \emph{Jubilee Issue of Hydrogen: A Fundamental System in All States}, Special issue of J. Phys. B \textbf{2018}.
\bibitem{Harrison2005} P. Harrison, \emph{Quantum Wells, Wires and Dots: Theoretical and Computational Physics of Semiconductors Nanostructure}, Wiley-Interscience, New York, 2005.
 \bibitem{Herschbach1993} D. R. Herschbach, J. Avery, O. Goscinski (Eds), \textit{Dimensional Scaling in Chemical Physics}, Kluwer Acad. Publ., London, 1993.
%
\bibitem{Tsipis1996} C.T. Tsipis, V.S. Popov, D. R. Herschbach and J. S. Avery (Eds) \emph{New Methods in Quantum Theory}, Kluwer Academic Publishers, Dordrecht, 1996.
%
\bibitem{Herschbach1996} D. R. Herschbach, Int. J. Quantum Chem., \textbf{57} (1996) 295 .
%
\bibitem{Herschbach2020} K.J.B. Ghosh, S. Kais, D. R. Herschbach, Unorthodox dimensional interpolations for He, Li, Be atoms and hydrogen molecule, Arxiv:2004.11489v1 [quant-ph] 23 Apr. 2020
\bibitem{Nieto1979} M. M. Nieto, Am. J. Phys., {\bf 47} (1979) 1067.

\bibitem{Witten2018} E. Witten, La Rivista del Nuovo Cimento \textbf{43} (2020) 187.

\bibitem{Toranzo2016} I. V. Toranzo, A. Mart\'{i}nez-Finkelshtein, J. S. Dehesa, J. Math. Phys., \textbf{57} (2016) 08219.
\bibitem{Toranzo2017b} D. Puertas-Centeno, N. Temme, I.V. Toranzo, J. S. Dehesa, J. Math. Phys., \textbf{58} (2017) 103302.

\bibitem{Puertas2018a} I. V. Toranzo, D. Puertas-Centeno, J. S. Dehesa, J. Stat. Mech.: Theory Exp. \textbf{2018} (2018) 073203.

\bibitem{Witten1980}  E. Witten, Phys. Today \textbf{33}, 38 (1980).

\bibitem{Hoof2016} G 't Hooft, The quantum black hole as a hydrogen atom: Microstates Without Strings Attached. 	ArXiv:1605.05119 [gr-qc] (2016)
\bibitem{Corda2018} C. Corda, F. Feleppa, The Quantum Black Hole as a Gravitational Hydrogen Atom. Preprints 2018, 2018100413 (doi: 10.20944/preprints201810.0413.v3).
%
\bibitem{Bures2015} M. Bures,  \emph{Quantum Physics With Extra Dimensions}, PhD thesis, Masaryk University, 2015.
\bibitem{Sen2012} K.D. Sen (Ed), {\it Statistical Complexity: Applications in Electronic Structure}, Springer, Berlin, 2012.
\bibitem{Lopez2013} S. L\'opez-Rosa, I.V. Toranzo, P. Sanchez-Moreno, J. S. Dehesa, J. Math. Physics   \textbf{54} (2013) 052109.
\bibitem{Sobrino-coll2017} N. Sobrino-Coll, D. Puertas-Centeno, I. V. Toranzo, J. S. Dehesa, J. Stat. Mech.: Theory and Exp.  \textbf{8} (2017) 083102.

\bibitem{Burgbacher1999} F. Burgbacher, C. Lammerzahl, A. Mac\'ias, J. Math. Phys. \textbf{40}  625 (1999).
\bibitem{Andrew1990} K. Andrew, J. Supplee, Am. J. Phys. \textbf{58} (1990) 1177.
\bibitem{Li2007} S. S. Li, J. B. Xia, Phys. Lett. A \textbf{366} (2007) 120.

\bibitem{Pathak1981} R. K. Pathak, S. R. Gadre, J. Chem. Phys. \textbf{74}  (1981) 5925.


\bibitem{Nagy1999} A. Nagy, S. Liu, R.G. Parr, \textit{Phys. Rev. A}  {\bf 59} (1999) 3349.
\bibitem{Liu1996} S. Liu, R.G. Parr, \textit{Phys. Rev. A}  {\bf 53} (1996) 2211.

\bibitem{Liu1997} S. Liu, R.G. Parr, Phys. Rev. A \textbf{55} (1997) 1792.
\bibitem{Angulo2000} J.C. Angulo, E. Romera, J.S. Dehesa, J. Math. Phys. \textbf{41} (2000) 7906.
\bibitem{Olver2010} F. W. J. Olver, D. W. Lozier, R. F. Boisvert, C. W. Clark (Eds), NIST Handbook of Mathematical Functions, Cambridge University Press, New York, 2010.

\bibitem{Fock1935} V.A. Fock, Z. Physik \textbf{98} (1935) 145.
\bibitem{Podolsky1929} B. Podolsky and L. Pauling, Phys. Rev. \textbf{34}  (1929) 109.
\bibitem{Hey1993} J.D. Hey, Am. J. Phys. \textbf{61} (1993) 28.

 \bibitem{Luke1969} Y. L. Luke, \textit{The Special Functions and their Approximations}, Vol. 2, Academic Press, New York, 1969.
\bibitem{Slater1966} L.J. Slater, \textit{Generalized Hypergeometric Functions}, Cambridge Univ. Press, Cambridge, 1966.

\bibitem{Drake1990} G.W.F. Drake, R.A. Swainson, \emph{Phys. Rev. A} {\bf 42} (1990) 1123.

 \bibitem{Andrae1997} D. Andrae, \emph{J. Phys. B: At. Mol. Opt. Phys.} {\bf 30} (1997) 4435.
 
 \bibitem{Tarasov2004} V.F. Tarasov, \emph{Int. J. Mod. Phys. B} {\bf 18} (2004) 3177.

 \bibitem{Vanassche2000} W. Van Assche, R. J. Y\'a\~nez, R. Gonz\'alez-F\'erez, J. S. Dehesa, J.Math. Phys. \textbf{41}  (2000) 6600.
  \bibitem{Ray1988} A. Ray, K. Mahata, P. P. Ray, Am. J. Phys. {\bf 56}, (1988) 462.
  \bibitem{Szymanski2020} T. Szymanski, J. K. Freericks, Algebraic derivation of Kramers-Pasternack relations based on the Schr\"{o}dinger factorization method,  arXiv:2007.11158v1 [quant-ph] 22 July 2020. 
  \bibitem{Hey1993b} J.D. Hey, Amer. J. Phys. \textbf{61} (1993) 741.

 \bibitem{Lewanowicz1985} S. Lewanowicz, Math Comput. \textbf{45} (1985) 521.
 \bibitem{Koepf2014} W. Koepf, \textit{Hypergeometric Summation. An Algorithmic Approach to Summation and Special Functions Identities}. 2nd. edition, Springer, London, 2014.
\bibitem{Koepf2020} W. Koepf, Computer Algebra, Power Series and Summation. In \textit{Orthogonal Polynomials}, in: M. Foupouagnigni, W. Koepf(Eds), AIMSVSW 2018. Tutorials, Schools, and Workshops in the Mathematical Sciences. Birkhauser, Cham., 2020.
\bibitem{Adkins} G.S. Adkins, M.F. Alam, C. Larison, R. Sun, Phys. Rev. A 101 (2020) 042511.

\bibitem{Buyarov1999} V.S. Buyarov, J.S. Dehesa, A. Martinez-Finklshtein, E. B. Saff, J. Approx. Theory \textbf{99} (1999) 153.

\bibitem{Aptekarev2010} A. I. Aptekarev, J.S. Dehesa, A. Martinez-Finkelshtein, R.J. Y\'{a}\~nez, \emph{J. Phys. A: Math. Theor.} 43  (2010) 145204.

\bibitem{Pathak1986} R.K. Pathak, B.S. Sharma and A.J.Thakkar, \textit{J. Chem. Phys.} {\bf 85} (1986) 958.


\bibitem{Heim1994} T.A. Heim, J. Phys. B: At. Mol. Opt. Phys. \textbf{27}  (1994).
\bibitem{Shiell2003} R. C. Shiell, Rydberg states, in: S. Wilson(Ed), \textit{Handbook of Molecular Physics and Quantum Chemistry},  John Wiley and Sons, New York, 2003. Part IV, pag. 423.

\bibitem{Costigliola2019} T. Maimbourg, J. C. Dyre, L. Costigliola, Density scaling of generalized Lennard-Jones fluids in different dimensions, ArXiv:1912.08176v1 [cond-mat.soft] 17 Dec 2019.

\bibitem{Daubechies1983} I. Daubechies, \textit{Comm. Math. Phys.} {\bf 90} (1983) 511.



\bibitem{Parr1989}  R. G. Parr and W. Yang, \textit{Density-Functional Theory of Atoms and Molecules}, Oxford University Press, Oxford, 1989.


\bibitem{Angulo1993}J. C. Angulo, J. Phys. A \textbf{26} (1993) 6493.

\bibitem{Angulo1994}J. C. Angulo, Phys. Rev. A \textbf{50} (1994) 311.

\bibitem{Zozor2011}S. Zozor, M. Portesi, P. S\'anchez-Moreno, J.S. Dehesa,  Phys. Rev. A 83 (2011) 052107

\bibitem{Folland1997} G. B. Folland, A. Sitaram, J. Fourier Anal. Appl. \textbf{3} (1997) 207. 
\bibitem{Angulo2011} J. C. Angulo, Phys. Rev. A \textbf{83} (2011) 062102.



\bibitem{Heisenberg1927} W. Heisenberg Z. Phys. 43 (1927) 172. Also in: J.A. Wheeler and W.H. Zurek(Eds), \textit{Quantum Theory and Measurement}, Princeton Univ. Press, Princeton, NJ, 1983, p. 62-84.
\bibitem{Kennard1927} E. H. Kennard, Z. Phys. \textbf{44} (1927) 326.
\bibitem{Sanchez2006} P. S\'anchez-Moreno, R. Gonz\'alez-F\'erez, J. S. Dehesa, New J. Phys. \textbf{8} (2006) 330.

\bibitem{Beckner1995} W. Beckner, Proc. Am. Math. Soc. \textbf{123} (1995) 1897.
\bibitem{Dehesa2007njp} J.S. Dehesa, R. Gonz\'alez-F\'erez, P. S\'anchez-Moreno, R. J. Y\'a\~nez,  New Journal of Physics \textbf{9} (2007) 131.
\bibitem{Toranzo2015} I.V. Toranzo, S. L\'opez-Rosa, R.O. Esquivel, J.S. Dehesa, \textit{Phys. Rev.} A {\bf 91} (2015) 062122.




\bibitem{Porras1990}  I. Porras, F.J. G\'alvez, \textit{Phys. Rev.} A {\bf 41} (1990) 4052.

\bibitem{Porras1995}  I. Porras, F.J. G\'alvez, \textit{Int. J. Quant. Chem.} {\bf 56} (1995) 763.
\bibitem{Lieb1983} E. H. Lieb, \textit{Int. J. Quant. Chem.} {\bf 24} (1983) 243.
\bibitem{Hundertmark2007} D. Hundertmark, Some bound states problems in Quantum Mechanics, in: F. Gesztesy, P. Deift, C. Galvez, P. Perry, W. Schlag (Eds), \textit{Spectral Theory and Mathematical Physics: A Festschrift in Honor of Barry Simon 60th Birthday },  \textit{Proc. Sympos. Pure Math.} {\bf 76} (2007) 463.
\bibitem{Lieb2010}  E. H. Lieb, R. Seiringer, \textit{The Stability of Matter in Quantum Mechanics}, Cambridge University Press, Cambridge, 2010.

\bibitem{Dehesa1989}  J.S. Dehesa, F.J. G\'alvez, I. Porras, \textit{Phys. Rev.} A {\bf 40} (1989) 35.


\bibitem{Galvez1987} F.J. G\'alvez, J.S. Dehesa, Phys. Rev. A {\bf 35}, (1987) 2384.
\bibitem{Dehesa1988} J. S. Dehesa, F.J. G\'alvez, Phys. Rev. A {\bf 37} (1988) 3634.
\bibitem{Toranzo2014} I.V. Toranzo, P. S\'anchez-Moreno, R.O. Esquivel, J.S. Dehesa,   \textit{Chem. Phys. Lett.}  {\bf 614} (2014) 1.
\bibitem{Toranzo2016jpa} I.V. Toranzo, S. L\'opez-Rosa, R. O. Esquivel, J. S. Dehesa, J. Phys. A: Math. Theor. \textbf{49}  (2016) 025301.
\bibitem{Angulo2012jmp} J.C. Angulo, P. A. Bouvrie, J. Antol\'in, J. Math. Phys. \textbf{53} (2012) 043512.


\bibitem{Guerrero2011} A. Guerrero, P. S\'anchez-Moreno, J. S. Dehesa,   \textit{Phys. Rev. A} \textbf{84} (2011) 042105.


\bibitem{Bialynicki2006} I. Bialynicki-Birula, Phys. Rev. A \textbf{74}, (2006) 052101.
\bibitem{Zozor2008} S. Zozor, M. Portesi, C. Vignat, Physica A \textbf{387} (2008) 4800.

\bibitem{Sanchez-Moreno2013} P. S\'anchez-Moreno, A. Zarzo, J. S. Dehesa, A. Guerrero, Appl. Math. Comput. \textbf{223} (2013) 25.

\bibitem{Srivastava1988} H. M. Srivastava, Astrophys. Space Sci., 150 (1988) 251.


\end{thebibliography}
\end{document}